\newcommand{\beq}{\begin{equation}}
\newcommand{\eeq}{\end{equation}}

\newcommand{\bear}{\begin{eqnarray}}
\newcommand{\eear}{\end{eqnarray}}

\newcommand{\tn}{\textnormal}
\documentclass{JINST}
\usepackage{graphicx}
\usepackage{amssymb}
\usepackage{array}
\title{Characterization of a medium size Xe/TMA TPC instrumented with microbulk Micromegas, using low-energy $\gamma$-rays}

\author{
The NEXT collaboration

V.~\'Alvarez,$^{a}$ F.I.G.M.~Borges,$^{d}$ S.~C\'arcel,$^{a}$ J.~Castel,$^{b,c}$ S.~Cebri\'an,$^{b,c}$ A.~Cervera,$^{a}$ C.A.N.~Conde,$^{d}$ T.~Dafni,$^{b,c}$ T.H.V.T.~Dias,$^{d}$ J.~D\'iaz,$^{a}$ M.~Egorov,$^{f}$ R.~Esteve,$^{g}$ P.~Evtoukhovitch,$^{h}$ L.M.P.~Fernandes,$^{d}$ P.~Ferrario,$^{a}$ A.L.~Ferreira,$^{i}$ E.D.C.~Freitas,$^{d}$ V.M.~Gehman,$^{f}$ A.~Gil,$^{a}$ A.~Goldschmidt,$^{f}$ H.~G\'omez,$^{b,c}$ J.J.~G\'omez-Cadenas,$^{a}$\thanks{Spokesperson (gomez@mail.cern.ch).}~ D. Gonz\'alez-D\'iaz,$^{b,c}$\thanks{Corresponding author.}~ R.M.~Guti\'errez,$^{j}$ J.~Hauptman,$^{k}$ J.A.~Hernando Morata,$^{l}$ D.C.~Herrera,$^{b,c}$ F.J.~Iguaz,$^{b,c}$ I.G.~Irastorza,$^{b,c}$ M.A.~Jinete,$^{j}$ L.~Labarga,$^{m}$ A.~Laing,$^{a}$  I.~Liubarsky,$^{a}$ J.A.M.~Lopes,$^{d}$ D.~Lorca,$^{a}$ M.~Losada,$^{j}$ G.~Luz\'on,$^{b,c}$ A.~Mar\'i,$^{g}$ J.~Mart\'in-Albo,$^{a}$ A.~Mart\'inez,$^{a}$ G. Martínez-Lema,$^{l}$ T.~Miller,$^{f}$ A.~Moiseenko,$^{h}$ F.~Monrabal,$^{a}$ C.M.B.~Monteiro,$^{d}$ F.J.~Mora,$^{g}$ L.M.~Moutinho,$^{i}$ J.~Mu\~noz Vidal,$^{a}$ H.~Natal da Luz,$^{d}$ G.~Navarro,$^{j}$ M.~Nebot-Guinot,$^{a}$ D.~Nygren,$^{f}$ C.A.B.~Oliveira,$^{f}$ R.~Palma,$^{n}$ J.~P\'erez,$^{o}$ J.L.~P\'erez Aparicio,$^{n}$ J.~Renner,$^{f}$ L.~Ripoll,$^{p}$ A.~Rodr\'iguez,$^{b,c}$ J.~Rodr\'iguez,$^{a}$ F.P.~Santos,$^{d}$ J.M.F.~dos Santos,$^{d}$ L.~Segui,$^{b,c}$ L.~Serra,$^{a}$ D.~Shuman,$^{f}$ A. Sim\'on,$^{a}$ C.~Sofka,$^{q}$ M.~Sorel,$^{a}$ J.F.~Toledo,$^{g}$ A.~Tom\'as,$^{b,c}$ J.~Torrent,$^{p}$ Z.~Tsamalaidze,$^{h}$ D.~V\'azquez,$^{l}$ J.F.C.A.~Veloso,$^{i}$ J.A.~Villar,$^{b,c}$ R.C.~Webb,$^{q}$ J.T.~White$^{q}$\thanks{Deceased.} and N.~Yahlali,$^{a}$
\newline
\newline
F.~Aznar,$^{b,c}$ D.~Calvet,$^{e}$ F.~Druillole,$^{e}$ E.~Ferrer-Ribas,$^{e}$ J.~A.~Garc\'ia,$^{b,c}$ I.~Giomataris,$^{e}$ J.~Gracia,$^{b,c}$ A.~Le Coguie,$^{e}$ J.P.~Mols,$^{e}$ P. Pons,$^{b}$ E. Ruiz$^{b}$

\\
\llap{$^{a}$}
Instituto de F\'isica Corpuscular (IFIC), CSIC \& Universitat de Val\`encia\\
Calle Catedr\'atico Jos\'e Beltr\'an, 2, 46980 Paterna, Valencia, Spain\\
\llap{$^{b}$}
Laboratorio de F\'isica Nuclear y Astropart\'iculas, Universidad de Zaragoza\\
Calle Pedro Cerbuna 12, 50009 Zaragoza, Spain\\
\llap{$^{c}$}
Laboratorio Subterr\'aneo de Canfranc\\
Paseo de los Ayerbe s/n, 22880 Canfranc Estación, Huesca, Spain\\
\llap{$^{d}$}
Departamento de Fisica, Universidade de Coimbra\\
Rua Larga, 3004-516 Coimbra, Portugal\\
\llap{$^{e}$}IRFU, Centre d'\'Etudes Nucl\'eaires de Saclay (CEA-Saclay)\\
91191 Gif-sur-Yvette, France\\
\llap{$^{f}$}
Lawrence Berkeley National Laboratory (LBNL)\\
1 Cyclotron Road, Berkeley, California 94720, USA\\
\llap{$^{g}$}
Instituto de Instrumentaci\'on para Imagen Molecular (I3M), Universitat Polit\`ecnica de Val\`encia\\
Camino de Vera, s/n, Edificio 8B, 46022 Valencia, Spain\\
\llap{$^{h}$}
Joint Institute for Nuclear Research (JINR)\\
Joliot-Curie 6, 141980 Dubna, Russia\\
\llap{$^{i}$}Institute of Nanostructures, Nanomodelling and Nanofabrication (i3N), Universidade de Aveiro\\
Campus de Santiago, 3810-193 Aveiro, Portugal\\
\llap{$^{j}$}
Centro de Investigaciones en Ciencias B\'asicas y Aplicadas, Universidad Antonio Nari\~no\\
Carretera 3 este No.\ 47A-15, Bogot\'a, Colombia\\
\llap{$^{k}$}
Department of Physics and Astronomy, Iowa State University\\
12 Physics Hall, Ames, Iowa 50011-3160, USA\\
\llap{$^{l}$}
Instituto Gallego de F\'isica de Altas Energ\'ias (IGFAE), Univ.\ de Santiago de Compostela\\
Campus sur, R\'ua Xos\'e Mar\'ia Su\'arez N\'u\~nez, s/n, 15782 Santiago de Compostela, Spain\\
\llap{$^{m}$}
Departamento de F\'isica Te\'orica, Universidad Aut\'onoma de Madrid\\
Campus de Cantoblanco, 28049 Madrid, Spain\\
\llap{$^{n}$}
Dpto.\ de Mec\'anica de Medios Continuos y Teor\'ia de Estructuras, Univ.\ Polit\`ecnica de Val\`encia\\
Camino de Vera, s/n, 46071 Valencia, Spain\\
\llap{$^{o}$}
Instituto de F\'isica Te\'orica (IFT), UAM/CSIC\\
Campus de Cantoblanco, 28049 Madrid, Spain\\
\llap{$^{p}$}
Escola Polit\`ecnica Superior, Universitat de Girona\\
Av.~Montilivi, s/n, 17071 Girona, Spain\\
\llap{$^{q}$}
Department of Physics and Astronomy, Texas A\&M University\\
College Station, Texas 77843-4242, USA\\

E-mail: \email{diegogon@unizar.es}

}

\abstract{NEXT-MM is a general-purpose high pressure (10\,bar, $\sim\!25$\,l active volume) Xenon-based TPC, read out in charge mode with an $0.8$\,cm$\times0.8$\,cm-segmented 700\,cm$^2$ plane (1152\,ch) of the latest microbulk-Micromegas technology. It has been recently commissioned at University of Zaragoza as part of the R\&D of the NEXT $0\nu\beta\beta$ experiment, although the experiment's first stage is currently being built based on a SiPM/PMT-readout concept relying on electroluminescence. Around 2 million events were collected during the last months, stemming from the low energy $\gamma$-rays emitted by a $^{241}$Am source when interacting with the Xenon gas ($E_\gamma= 26$, $30$, $59.5$\,keV). The localized nature of such events above atmospheric pressure, the long drift times, as well as the possibility to determine their production time from the associated $\alpha$ particle in coincidence, allow the extraction of primordial properties of the TPC filling gas, namely the drift velocity, diffusion and attachment coefficients. In this work we focus on the little explored combination of Xe and trimethylamine (TMA) for which, in particular, such properties are largely unknown. This gas mixture offers potential advantages over pure Xenon when aimed at Rare Event Searches, mainly due to its Penning characteristics, wave-length shifting properties and reduced diffusion, and it is being actively investigated by our collaboration. The chamber is currently operated at 2.7\,bar, as an intermediate step towards the envisaged 10\,bar. We report here its performance as well as a first implementation of the calibration procedures that have allowed the extension of the previously reported energy resolution to the whole readout plane ($10.6\%$\,FWHM@30\,keV).}


\begin{document}

\section{Introduction}\label{Introduction}


NEXT-100 is a $0\nu\beta\beta$ experiment to be located at the Laboratorio Subterr\'{a}neo de Canfranc (LSC) in the Spanish Pyrenees \cite{refCDR,refTDR}, with its first stage currently undergoing construction \cite{DGD}. Notably, it is the only next generation $0\nu\beta\beta$ experiment that uses a $\beta\beta$-isotope in gaseous phase ($^{136}$Xe, 100-150\,kg) and thus it aims at maximally exploiting the topological features of the $2e^{-}$ $\beta\beta$-decay mode \cite{NEXT_DEMO2}. The ability to recognize and separate the golden 2-blob signal from spurious (single-blob) $\gamma$ backgrounds is one of the experiment's key features, relying for that on a Time Projection Chamber (TPC) architecture. Its second strength, as compared to earlier approaches \cite{Luscher:1998sd}, is the use of proportional light multiplication (namely, electroluminescence/EL), a technique capable of providing near-intrinsic energy resolution in pure Xenon, down to 0.5\%\,FWHM$@Q_{\beta\beta}$ \cite{NEXT_LBNL}. The design goals as of the technical design report released by the collaboration \cite{refTDR} have been confirmed for 1\,kg-scale demonstrators in a series of papers \cite{NEXT_DEMO2, NEXT_LBNL, SiPM_Nadia, NEXT_DEMO}.

The fact that Xenon is in gas phase at standard $P$, $T$ conditions reduces the cost of the enrichment process (the isotopic content of $^{136}$Xe in natural Xenon is 9\%), making it one of the $\beta\beta$-isotopes with the lowest price/mass ratio. However the gas phase offers yet additional advantages, for instance allowing readily for the combination with suitable gas dopants. Admixtures of some $\%$ (sometimes sub-$\%$), specially in noble gases, are known to intensely modify the characteristics of the main gas by quenching or wavelength-shifting its scintillation spectrum \cite{Suzuki}, enhancing energy-transfer ionizing (Penning) reactions \cite{Sahin}, modifying the charge recombination process \cite{Ramsey} or affecting generic properties of the electron swarm like diffusion and drift velocity \cite{EL_port}, or attachment \cite{Attach}. A remarkable example currently under study, the trimethylamine molecule (TMA), potentially exhibits a desirable behaviour concerning all the above aspects when teaming up with Xenon\cite{Nygren}, such mixtures being however poorly known \cite{RamseyTMA, Diana1, Diana2}.
As an example, despite the electron drift velocity for pure Xenon (see \cite{NEXT_LBNL} for instance) and pure TMA \cite{TMA_drift} being long measured, data for Xe/TMA mixtures were not available until recently \cite{Diana2} to the best of the authors' knowledge.

We make use in this work of the recently commissioned NEXT-MM TPC \cite{Hector_MM} that, thanks to its large drift distance (38\,cm) and finely segmented charge readout, allows a more exhaustive characterization of Xe/TMA mixtures as compared to \cite{Diana2}, thus providing the diffusion and attachment coefficients.
On the other hand, and perhaps most notably, NEXT-MM currently features the largest pixelated readout plane world-wide (700\,cm$^2$) based on the novel microbulk-Micromegas technology \cite{Iguaz}. Hence, except for the pioneering work realized on $10$\,cm$^2$-readout TPCs in \cite{Diana1}, nothing is known about the performance of large systems based on this kind of readouts when operated in high pressure Xenon mixtures. In order to further scrutinize this latter aspect, as well as to properly substantiate some of the analysis procedures here described, this contribution is structured as follows: in section \ref{TPC cali} the experimental setup, the data taking and calibration procedures are introduced, together with a description of the main TPC performance in the pressure range 1-2.7\,bar; in section \ref{swarm} the extraction of some relevant parameters of the electron swarm for Xe/TMA is performed and a short discussion follows.


\section{Overall TPC behaviour and performance of the Micromegas readout at 1-2.7\,bar}\label{TPC cali}

\subsection{Description of the experimental setup}

\begin{figure}[h]
 \centering
 \includegraphics*[width=14cm]{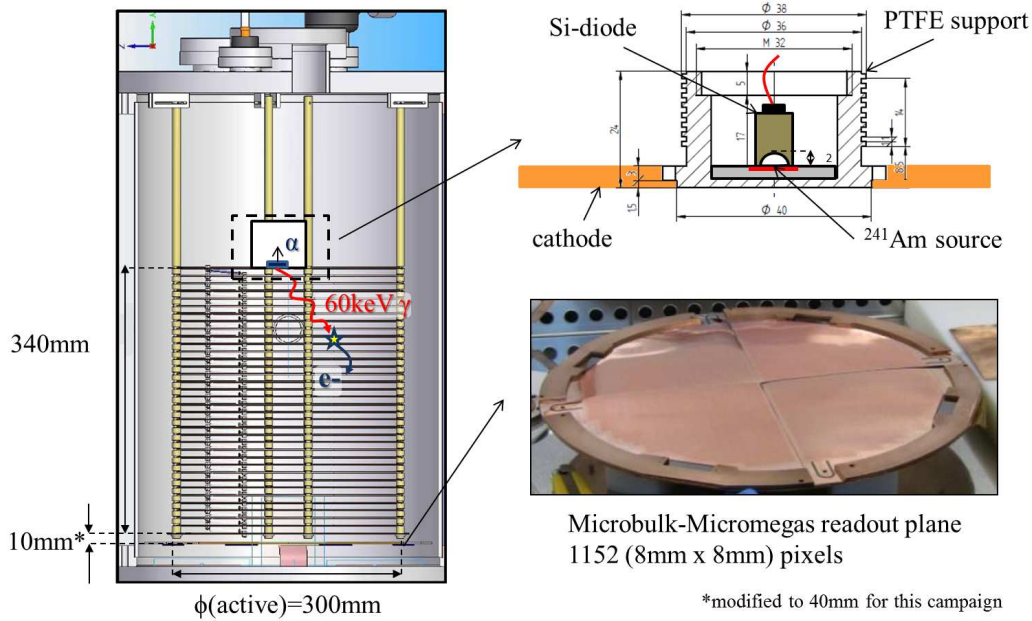}
 \caption{Experimental setup used for the characterization of NEXT-MM with low energy $\gamma$ photons, under a Xe/TMA mixture. Left: transverse section of NEXT-MM showing the field cage, source position and readout plane. Right-up: close-up of the $^{241}$Am source in its PTFE enclosure. Right-down: photo of the readout plane, showing the arrangement in 4 separate quadrants.}
 \label{layout}
 \end{figure}

The experimental setup used is very similar to the one described in \cite{Hector_MM}, that is shown in Fig. \ref{layout}. A radioactive source consisting of a thin layer of $^{241}$Am ($\mathcal{A}=500$\,Bq) electro-deposited in one of the two large faces of a cylindrical metal disk (2.5\,mm-thick) was employed. The disk itself was opaque to $\alpha$'s. It was enclosed in a PTFE structure attached to the TPC cathode (Fig.\ref{layout}, up-right) with its radioactive deposit facing the cathode of a silicon diode. The signal produced by an $\alpha$ particle traversing the diode ($\varepsilon_\alpha = 5.5$\,MeV) provided the start time ($T_0$) of the event, thus tagging the nuclear decay. Due to the limited solid angle coverage of the (extended) source, the diode rate reached only $r_{T_0}\sim40$\,Hz. For these measurements we made use of the fact that $^{241}$Am $\alpha$ particles are emitted largely in coincidence with 59.5\,keV $\gamma$'s from the $^{237}$Np daughter nuclei, that were able to enter and ionize the active volume of the TPC for about a fraction $f=1/15$ of all events. The relatively large mean free path of this emission ($\lambda_{59.5,Xe}=20$\,cm@1\,bar, table \ref{sources}) provided a sizeable ionization probability over the full TPC volume.

\begin{figure}[h]
\centering
\includegraphics*[width=\linewidth]{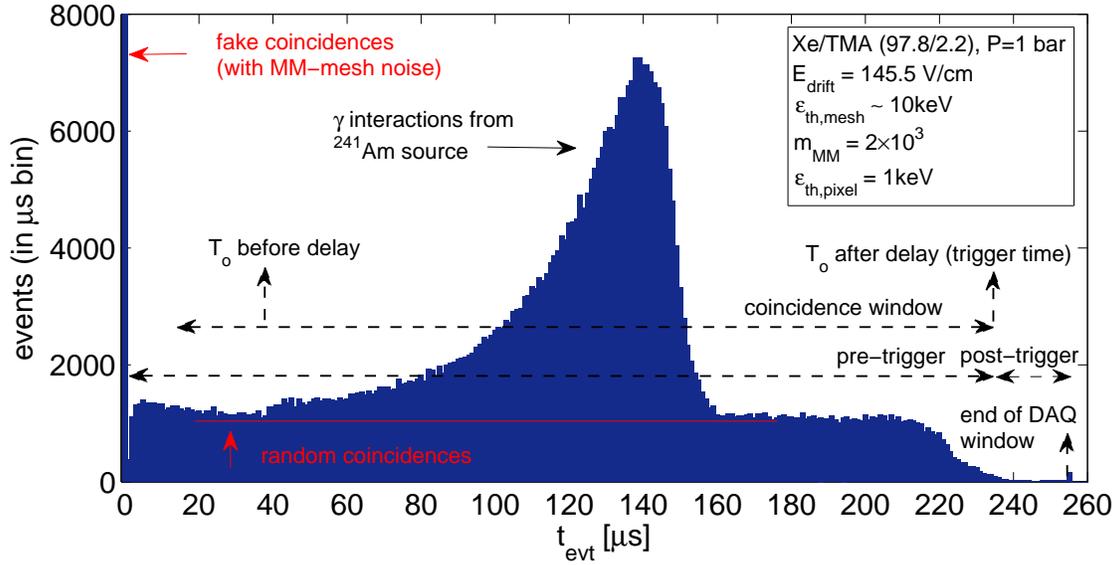}
\caption{Distribution of event times registered at NEXT-MM when triggering in coincidence with $\alpha$'s from a $^{241}$Am source, detected in an auxiliary Si-diode. The observed exponential shape is dominated by the mean free path of the $59.5$\,keV $\gamma$ emitted in the decay of $^{237}$Np to its ground state, modulated by solid angle and other geometrical effects. The edges of the TPC cathode and the readout plane are clearly discernible over a flat background. The chosen settings reflect typical working conditions described in the text.}
\label{tref}
\end{figure}

The trigger signal was built with the $T_0$-signal from the diode in coincidence with a positive signal from the TPC itself, in order to enhance the $\gamma$ content in data. The latter was formed by an `OR' of the signals induced at the cathode of each Micromegas quadrant (hereafter referred to as `mesh').
Application of an energy threshold of $10$\,keV on the mesh signals required operating the Micromegas plane with a gain around 2000, close to the spark limit previously reported in \cite{Diana1}.
The trigger rate $r_{trigger} = f \times r_{T_0} \simeq 2$\,Hz amounted to one  million events for each regular week of data taken, out of which 10\% were fake noise triggers and nearly 30\% random, typically (Fig. \ref{tref}).

For each triggered event the signals from the segmented anode plane of the Micromegas were amplified, sampled and stored with the help of a data acquisition system (DAQ) based on the one developed for the T2K experiment \cite{AFTER_T2K}, whose front end card (FEC) relies on the AFTER chip. This electronics, intrinsically capable of providing an accurate signal sampling down to 20\,ns time bins and low (sub-1000e- ENC) noise figure, is at the core of the NEXT-MM TPC concept. In the present experimental conditions it was convenient to select a time sampling in the range $0.3$-$0.9\,{\mu}$s as well as the largest shaping time available ($\tau=2\mu$s). The lowest amplification of the FEC was chosen, providing a dynamic range of 600\,fC.
Recorded events are shown later (Fig. \ref{PSA_all}) and several examples of extended tracks can be found in \cite{Hector_MM}. An absolute time can be associated to each event, defined as the mean of the pulses' peaking times ($t_{peak}$) obtained from all pulses ($N_{fired}$) above a given threshold ($\epsilon_{th,pixel}$):
\beq
t_{evt} =\frac{1}{N_{fired}}\sum_{i=1} ^{N_{fired}} t_{peak,i} \label{tarrival}
\eeq

The synchronization of the acquisition system with the arrival times of the $\gamma$-ray charge deposits to the readout plane ($t_{evt}$, from eq. \ref{tarrival}) can be illustrated with the help of Fig. 2, resorting for that to a raw sample of $5\times10^5$ events. The procedure can be summarized as follows. First, each mesh signal was amplified through a Canberra chain (models 2004 and 2022) and later discriminated. An `OR' signal was produced by a coincidence module (CAEN-N455) and its output enlarged with a dual timer (CAEN-2255B) to a width comfortably exceeding the anticipated drift time region. This procedure set the coincidence window, to which the $T_0$ diode signal (200\,ns width) was incorporated under an `AND' condition. In order for the resulting trigger signal to convey the $T_0$ on its rising edge, the diode signal had been previously delayed by a time exceeding the drift region but smaller than the coincidence window. At last, the window of the data acquisition system was opened upon the trigger arrival, by defining a certain amount of pre and post-trigger time, with the pre-trigger time chosen to exceed the duration of the coincidence window. The main merit of this experimental configuration is to allow for the accurate identification of the positions of the TPC cathode and the readout plane without resorting to external information. They appear as the edges of a nearly exponential distribution superimposed on a flat background, with a large fraction of coincidences coming from the source itself (after background subtraction the edge corresponding to the readout plane, situated around the arrow labeled `$T_o$ before delay' in Fig.\ref{tref}, can be further enhanced as shown in Fig. \ref{W_t_corr}).


A detailed description of the gas system can be found in \cite{Diana2}. For the present measurements Xenon (purity grade 6) and trimethylamine were used. Special care was taken with the latter, supplied in a bottle for which a large O$_2$ contamination was determined before hand with the help of a mass spectrometer. Purification was realized by first cryo-pumping the gas into a 2\,l stainless-steel bottle immersed in a Dewar flask filled with liquid N$_2$. After been reclaimed, the residual vapor in the bottle was pumped (still at cryogenic temperature) under the assumption that the O$_2$ present in the system remained in gas phase or with a sizeable vapor pressure. Whenever TMA was injected in the gas system this purification procedure was followed, and iterated if evidence was found of any O$_2$ contamination remaining.
In case the chamber had been previously opened in air (for maintenance or reparation works) a single bake-out cycle as described in \cite{Hector_MM} followed. In order to further enhance the quality of the gas, recirculation in close-loop through a FaciliTorr filter by SAES was performed at around 10\,Nl/h, and about 1-2 days were awaited before data taking started. According to the provider the filter can be operated under TMA, however an unwelcome reactive behaviour was observed reaching stationary conditions only after several minutes  \cite{Diana1}. Due to the large gas volume employed in these measurements, no sizeable effect could be noticed.

\begin{table}[h]
  \centering
  \begin{tabular}{|c|c|c|c|c|c|c|}
     \hline
     campaign & $\#$ runs & evts & $E_{drift}/P$\,[V/cm/bar] & $E_{MM}/P$\,[kV/cm/bar] & $P$\,[bar] & Xe/TMA \\
     \hline
     I & 2 & $8\times10^5$ & 145.5 & 54 & 1.0 & 97.8/2.2\\
     II & 5 & $10^6$ & $66.6$-$164.0$ & 54 & 1.0 & 97.8/2.2\\
     III  & 2 & $6\times10^5$ & 103.6 & 25 & 2.7 & 97.6/2.4\\
     \hline
   \end{tabular}
  \caption{Summary of the experimental campaigns conveyed in this work. The reduced field in the multiplication region ($E_{MM}$) is estimated through division of the voltage at the Micromegas cathode by the amplification gap, normalized to the working pressure. It corresponds to a gain of $\times 2000$ for the 1\,bar campaigns and $\times 1600$ for the 2.7\,bar one.}\label{runs}
\end{table}

\subsection{Data taking and analysis}

\begin{figure}[h]
 \centering
 \includegraphics*[width=7.5cm]{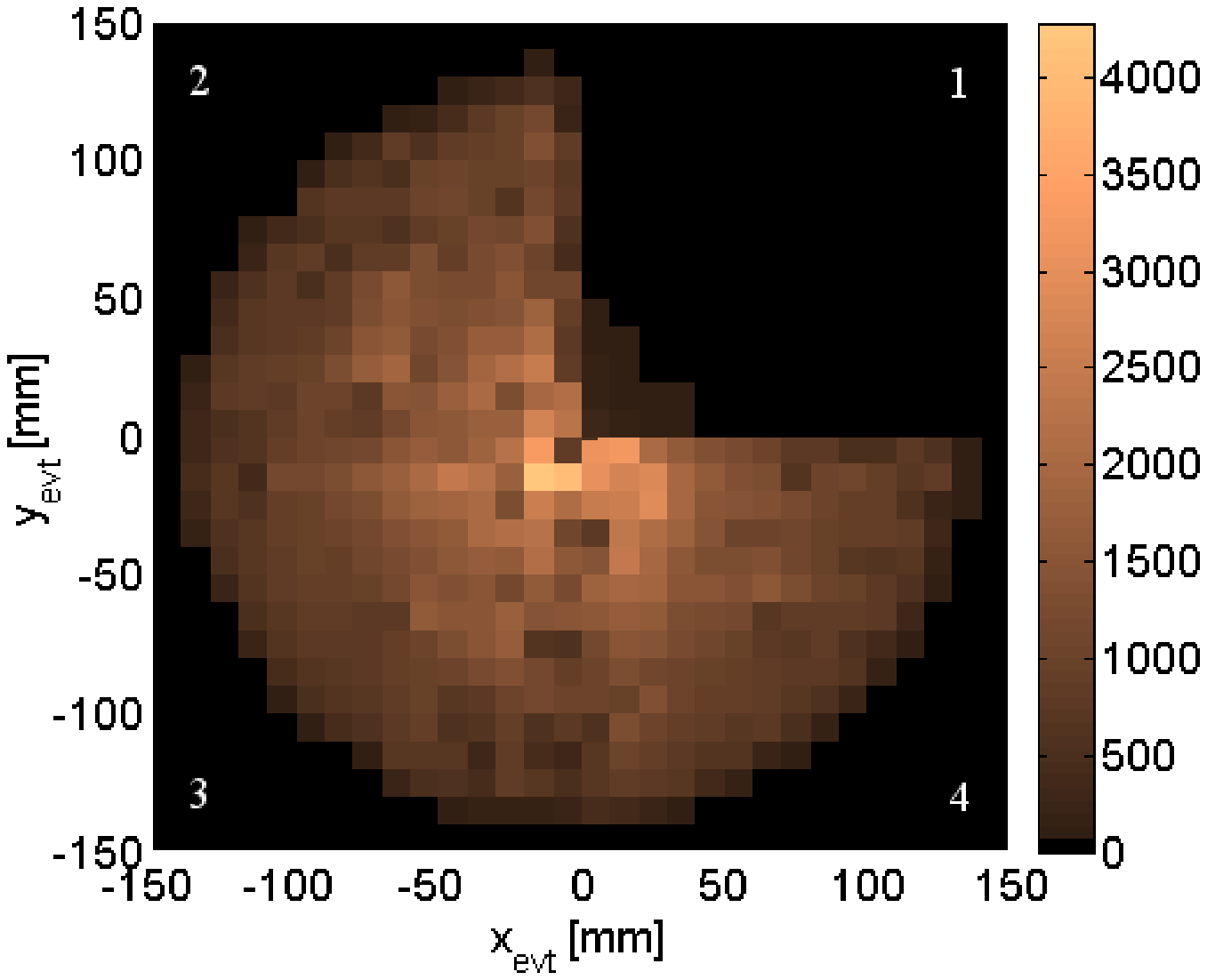}
 \includegraphics*[width=7.5cm]{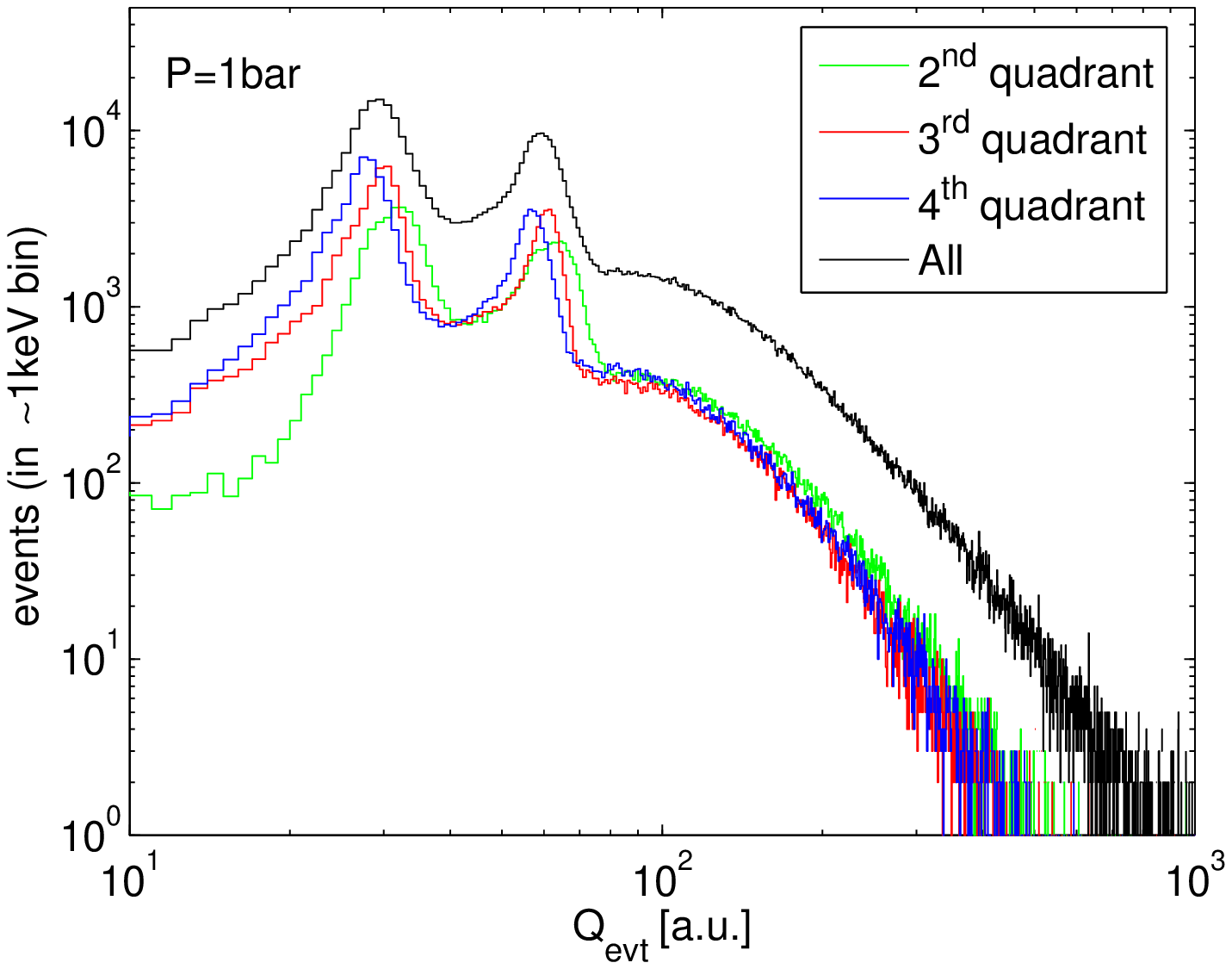}

 \includegraphics*[width=7.5cm]{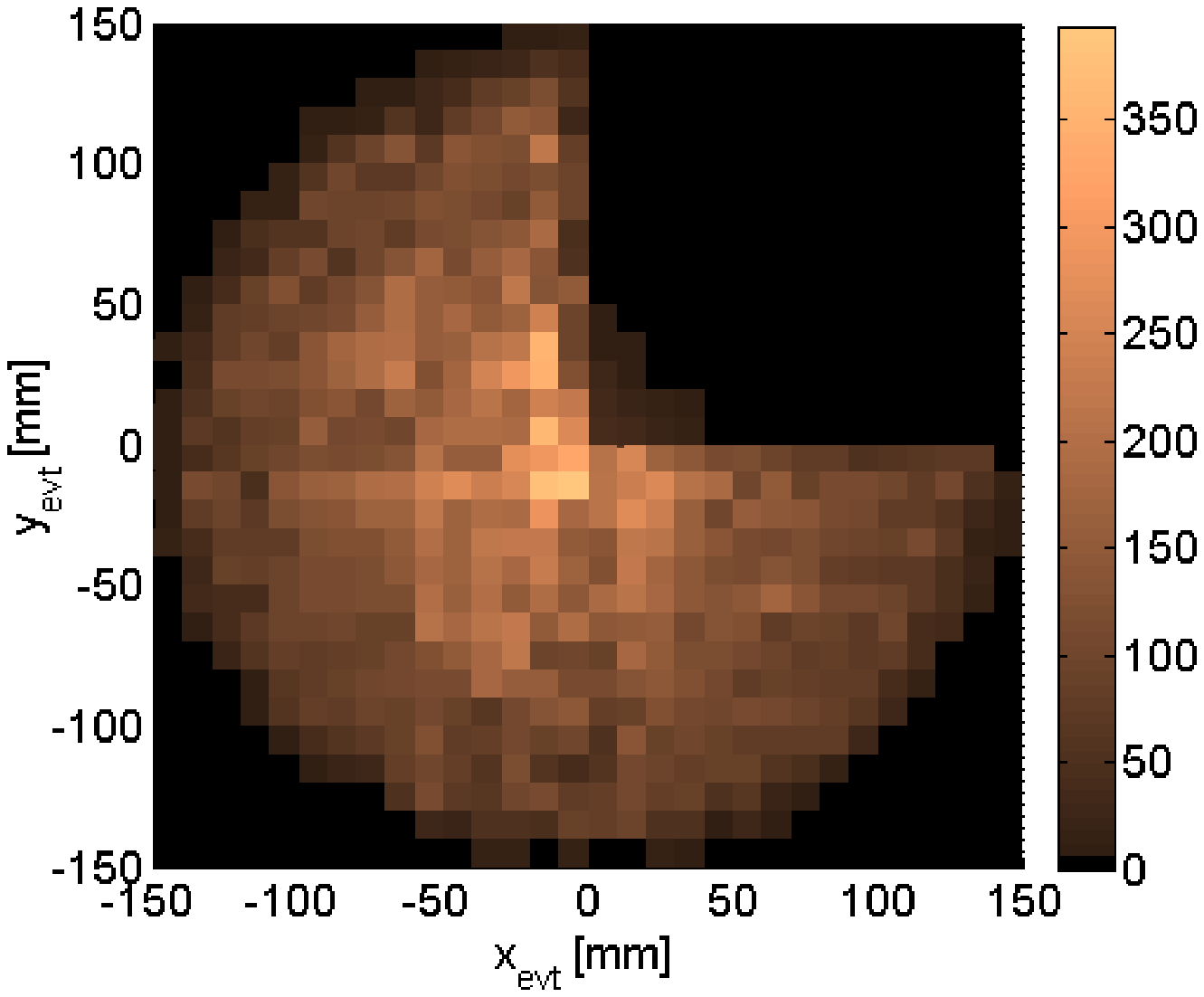}
 \includegraphics*[width=7.5cm]{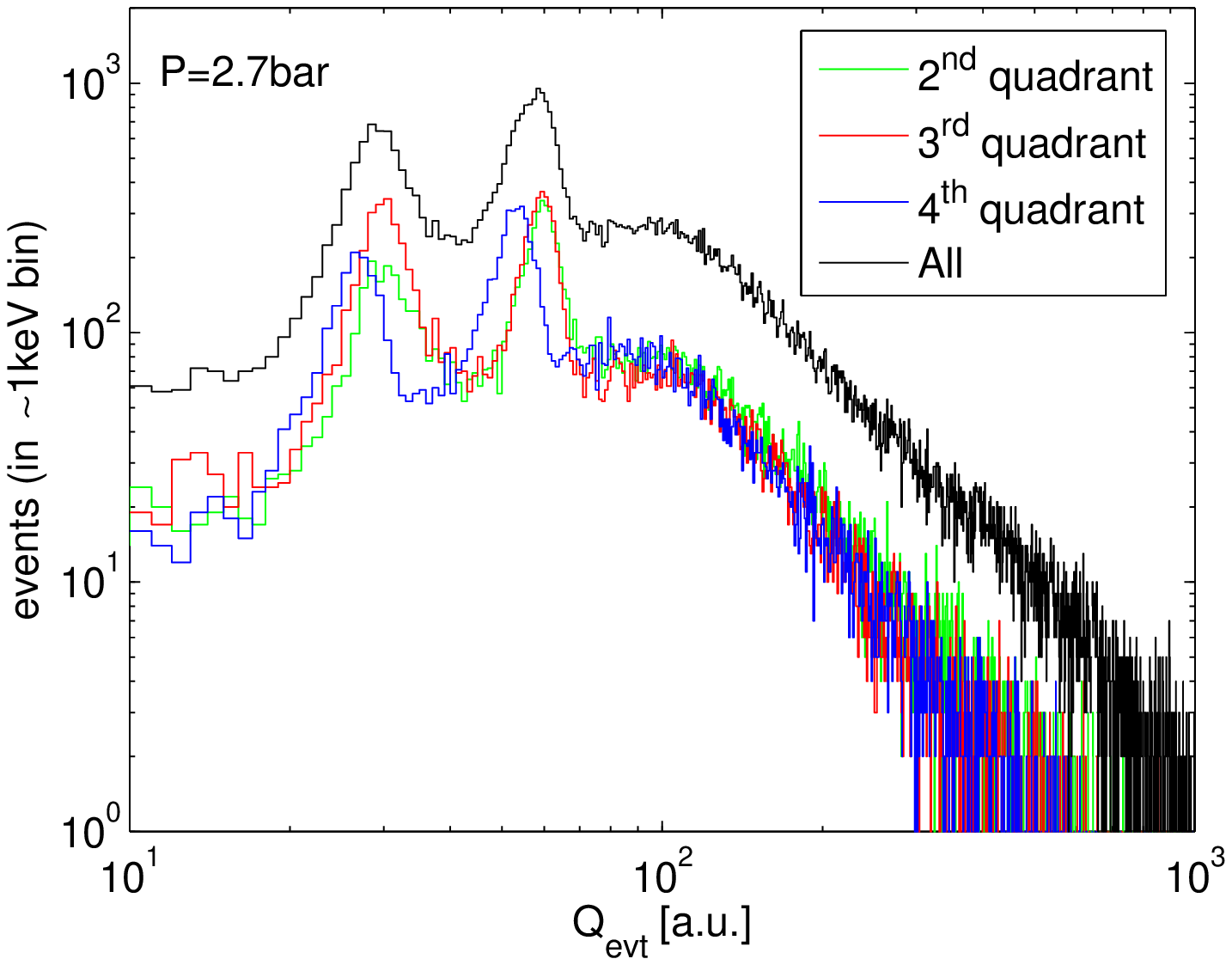}
 \caption{Left: $x$-$y$ transverse position distribution of $\gamma$ events from a $^{241}$Am source,
 as obtained at the TPC readout plane under a 1\,cm$\times 1$\,cm binning. Right: raw charge spectra with arbitrary normalization. The first row presents results obtained at 1\,bar ($m=2000$) and the second one at 2.7\,bar ($m=1600$).}
 \label{examples_P}
 \end{figure}

\begin{figure}[h]
 \centering
 \includegraphics*[width=14cm]{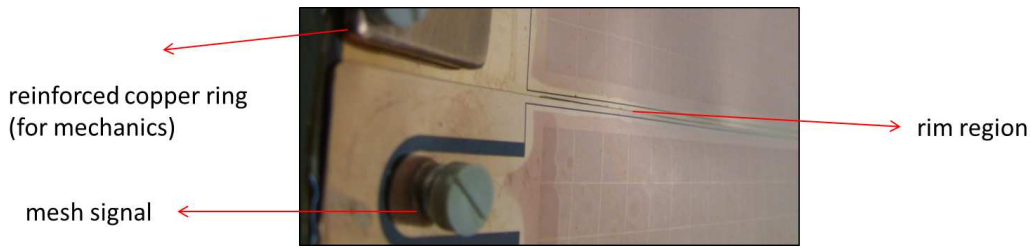}
 \caption{Close up of the `rim technique' used for reducing charge loss in multi-wafer Micromegas assemblies when aimed at large area coverage.}
 \label{rim}
 \end{figure}

Around $2\times10^6$ events were stored during 3 experimental campaigns, summarized in table \ref{runs}:
i) a high statistics campaign and ii) a drift-field scan, both at $P=1$\,bar, and iii) a high statistics campaign at $P=2.7$\,bar. For all three an internal insulation problem in the anode signal distribution of the first quadrant limited the TPC readout to $3/4$ of its total area, out of which 92\% was fully operational (for details see \cite{Hector_MM}).
The data analysis described in this section is based on simple algorithms and no detailed pulse shape analysis (PSA) has been attempted. It starts from the DAQ-stored 511-point waveforms of all $288\times3$ active electronic channels/pixels containing the raw information to be analyzed. Upon pedestal subtraction, a threshold in amplitude corresponding to $\sim$1\,keV was set for each channel. When crossing the threshold, the initial estimate of the pulse charge (obtained from the pulse amplitude) was refined through the determination of the pulse area, in order to correct for any residual ballistic deficit, providing $Q_i$. The position of the event in the transverse plane, $x_{evt}$, $y_{evt}$, could be then obtained through a standard linear weighting procedure based on the estimated charge:
\bear
&& Q_{evt}= \sum_i^{N_{fired}} Q_{i} \\
&& x_{evt}= \frac{1}{Q_{evt}} \sum_i^{N_{fired}} Q_{i} x_{i} \label{x_evt} \\
&& y_{evt}= \frac{1}{Q_{evt}} \sum_i^{N_{fired}} Q_{i} y_{i} \label{y_evt}
\eear
The raw charge distribution as well as the corresponding $x$-$y$ distribution of the triggered events are shown in Fig. \ref{examples_P}, for two of the runs belonging to the 1\,bar and 2.7\,bar campaigns (table \ref{runs}). The presence of the source can be clearly noticed at the chamber center.
Interestingly, there is no visible discontinuity in the region between the quadrants and the radial coverage reaches nearly the end of the readout plane: $R=15\,$cm (the inner radius of the field cage is 14\,cm, creating a $\sim\!1$\,cm shadow). The large homogeneity observed is largely due to the fact that a narrow `rim' region (Fig. \ref{rim}), surrounding each Micromegas quadrant was foreseen during the manufacturing process. This region, inter-connected for all quadrants, was biased $20\,V$ above the mesh voltage, reducing in this way the charge loss at the readout plane due to fringe fields and dead regions.

The charge distributions of Fig. \ref{examples_P} contain two prominent peaks on a feature-less background stemming from natural radioactivity. They can be interpreted with the help of table \ref{sources} in the following way: $\alpha$-decays from $^{241}$Am populate in overwhelming majority the $5/2^-$ excited state of $^{237}$Np, that decays to the ground state by emitting a photon of energy $\varepsilon_1=59.54$\,keV in 35.9\% of the cases \cite{Decay}. With smaller probability a $\varepsilon_2=26.3$\,keV photon can be emitted (2.3\%), while other emissions are at the sub-0.1\% level. According to \cite{Aprile} low-energy $\gamma$ interactions in Xenon take place predominantly with the atomic $K$-shell and so the resulting photo-electron will be accompanied by characteristic Xe X-ray emission ($K_{\alpha}$, $K_{\beta}$) in nearly 80\% of the cases. These photons have mean free paths of the order of 20\,cm at 1\,bar, thus having a sizeable probability to escape from the TPC active volume, leaving an escape peak energy of $\varepsilon_1\!-\!K_{\alpha,\beta}$. The reduction, as a function of pressure, of the low-energy peak content in Fig. \ref{examples_P} relative to the high-energy one is strongly in favour of this interpretation.\footnote{If both peaks were dominantly produced by photons emitted at the source the opposite behaviour as a function of pressure would be slightly favoured according to table \ref{sources}.} After calibration (Fig. \ref{Cal_plots}) 3 peaks can be indeed resolved in the 30\,keV region as anticipated from table \ref{sources}.

It is to be noted that the raw energy resolution of the readout plane as extracted from the 30\,keV peak is somewhere around 25\%\,FWHM. This number, a weak result for the standards of micro-pattern gaseous detectors, must be interpreted on the light of the effects that are still present at this level of data processing: gain variations between quadrants and within the quadrants, lack of event containment close to the borders and fringe fields contribute to the spread in the recorded charge, and are considered in some detail in the next sub-section.

\begin{table}[h]
  \centering
  \begin{tabular}{|c|c|c|c|c|}
     \hline
     type & energy[keV]& number/decay & $\lambda_{\gamma,Xe}@$1bar[cm] & $R_{e,Xe(csda)}@$1bar[cm]\\
     \hline
     main $^{237}$Np $\gamma$ & 59.54 & 0.359 & 22 & 2\\
     Xe-$K_{\beta}$ & 33.64 & - & 26 & 0.8\\
     Xe-$K_{\alpha}$& 29.80 & - & 19& 0.6\\
     escaped Xe-$K_{\alpha}$ & 29.74 & - & 19 & 0.6\\
     sec. $^{237}$Np $\gamma$ & 26.35 & 0.0231 & 14 & 0.45\\
     escaped Xe-$K_{\beta}$& 25.90 & - & 14 & 0.45 \\
     \hline
   \end{tabular}
  \caption{Characteristics of the $\gamma$ content of the $^{241}$Am decay (only emission probabilities above $10^{-3}$/decay are shown) \cite{Decay}, together with some $\gamma$ and $e^-$ properties from NIST\cite{NIST}. The atomic properties of Xenon are taken from \cite{Aprile}.}\label{sources}
\end{table}

\begin{figure}[h]
 \centering
 \includegraphics*[width=\linewidth]{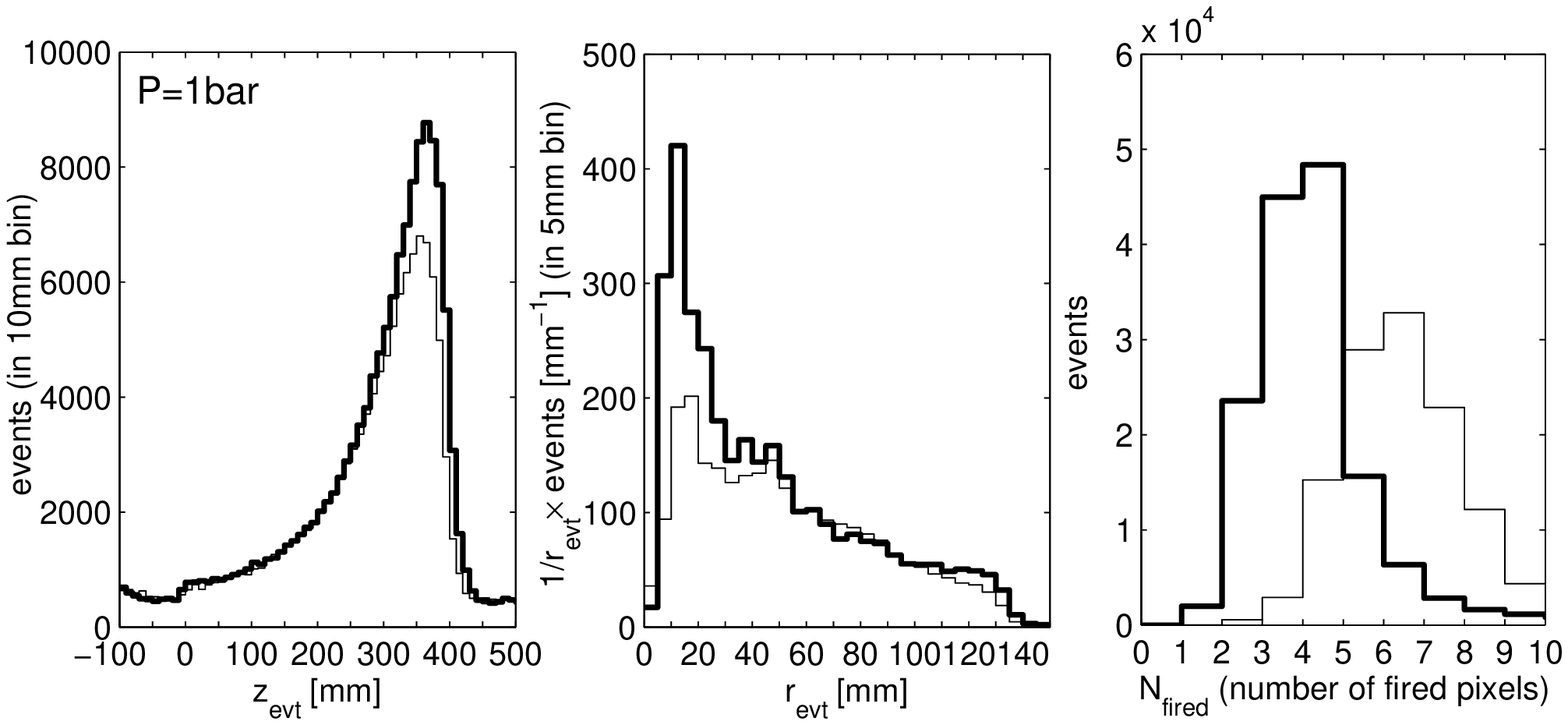}

 \includegraphics*[width=\linewidth]{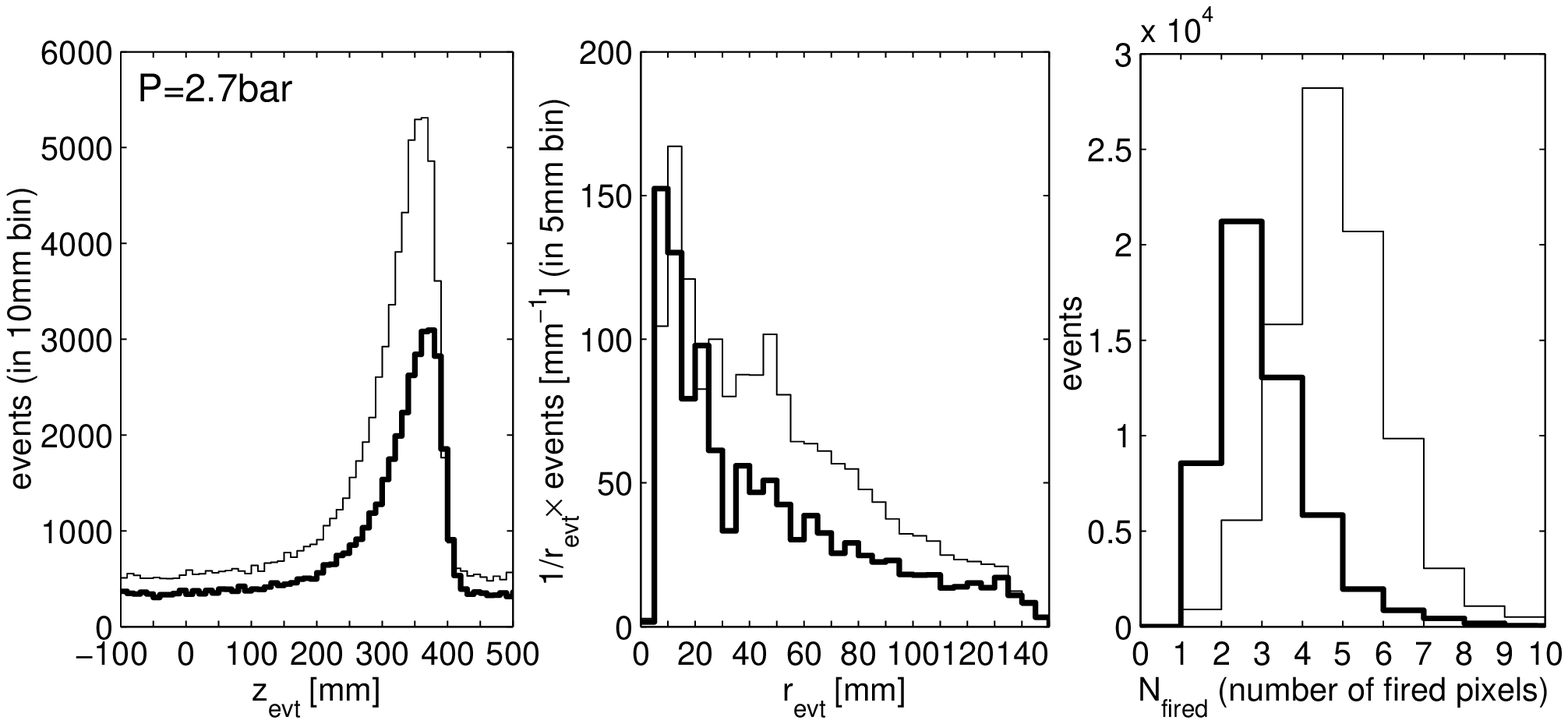}
 \caption{Spatial characteristics of the ionization created in the TPC by $\gamma$-rays produced by a $^{241}$Am source placed at the TPC cathode ($D=38$\,cm). Two energy regions have been selected after calibration ($20\,$keV$<\varepsilon_{evt}<40$\,keV and $50\,$keV$<\varepsilon_{evt}<70$\,keV) corresponding to the $\sim30\,$keV Xenon escape peak (thick line) and the full-absorption $59.5$\,keV one (thin line). Left: distribution along the drift region. Middle: radial distribution. Right: pixel multiplicity.}
 \label{examples_P_1D}
 \end{figure}

The $x$-$y$ distribution in Fig. \ref{examples_P} is blurred by instrumental and statistical effects and by the physical properties of the $\gamma$ and background tracks. For a better characterization of the $\gamma$ content, of interest in this work, Fig. \ref{examples_P_1D} shows 1D-distributions gated around the two main peaks present in the raw charge spectrum: $\sim\!30$\,keV (thick line) and $59.5$\,keV (thin line). For representation we make use of the drift velocity ($v_d$) obtained in section \ref{swarm}, that allows converting the measured time to the position along the drift ($z$) direction:
\beq
z_{evt} = v_d \times (t_{evt}-t_{ano}) \label{zevt}
\eeq
where $t_{ano}$ is the time corresponding to the left-edge of the event time distribution (Fig. \ref{tref}), interpreted as the time of an event produced right at the Micromegas plane. The radial position of the ionization cloud is defined from the pixelization as:
\beq
r_{evt} = \sqrt{x^2_{evt} + y^2_{evt}} \label{r_evt}
\eeq
Fig. \ref{examples_P_1D}-left shows the $z$-distribution obtained from eq. \ref{zevt} for two different pressures, with the readout plane situated at $z=0$\,cm and the TPC cathode ($z=D=38$\,cm) near the maximum of the distribution. It is apparent the drastic reduction of the 30\,keV peak content as compared to the 59.5\,keV one with increasing pressure due to the reduced escape probability. The pressure increase leads also to the shrinking of the $z$-distribution and to the relative increase of random coincidences in the region close to the readout plane, that is barely discernible. The radial distributions (Fig. \ref{examples_P_1D}-middle) show a decreasing character although not so marked ($R\lesssim\lambda_{59.5,Xe}<D$). The pixelization of the space and the non-perfect connectivity (\cite{Hector_MM}) introduce a more irregular pattern on them. Although the slope of the $30$\,keV and $59.5$\,keV space distributions could be expected to be the same (both charge peaks proceed from the interaction of a primary $59.5$\,keV photon), the high probability of characteristic Xe X-ray emission for $59.5$\,keV interactions modifies the picture. Since this secondary photon emission often results in a relatively distant second ionization cluster ($\lambda_{K_{\alpha, \beta}}=19$-$26$\,cm), the reconstructed position is artificially averaged out causing a smoother behaviour than for the single-cluster escape peak. The average pixel multiplicity for each ionization cluster (Fig. \ref{examples_P_1D}-right) is seen to be around 3-4 (1\,bar) and 2 (2.7\,bar) and its strong reduction with pressure is indicative of the fact that cluster sizes stem from the size of the ionization cloud and transverse diffusion, with instrumental effects (cross-talk or charge sharing) having a minor role.

\subsection{Gain calibration and system behavior}

\begin{figure}[h]
\centering
\includegraphics*[width=4.5cm]{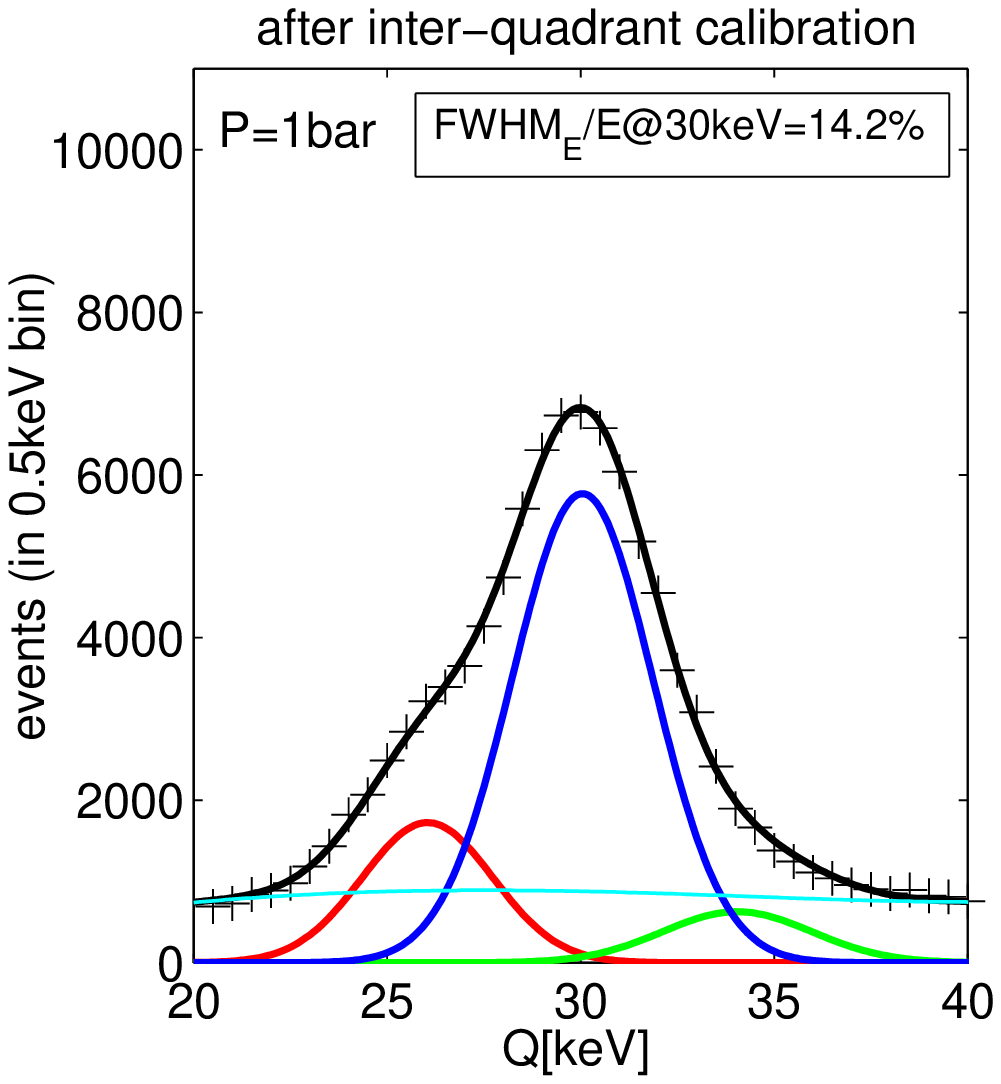}
\includegraphics*[width=4.5cm]{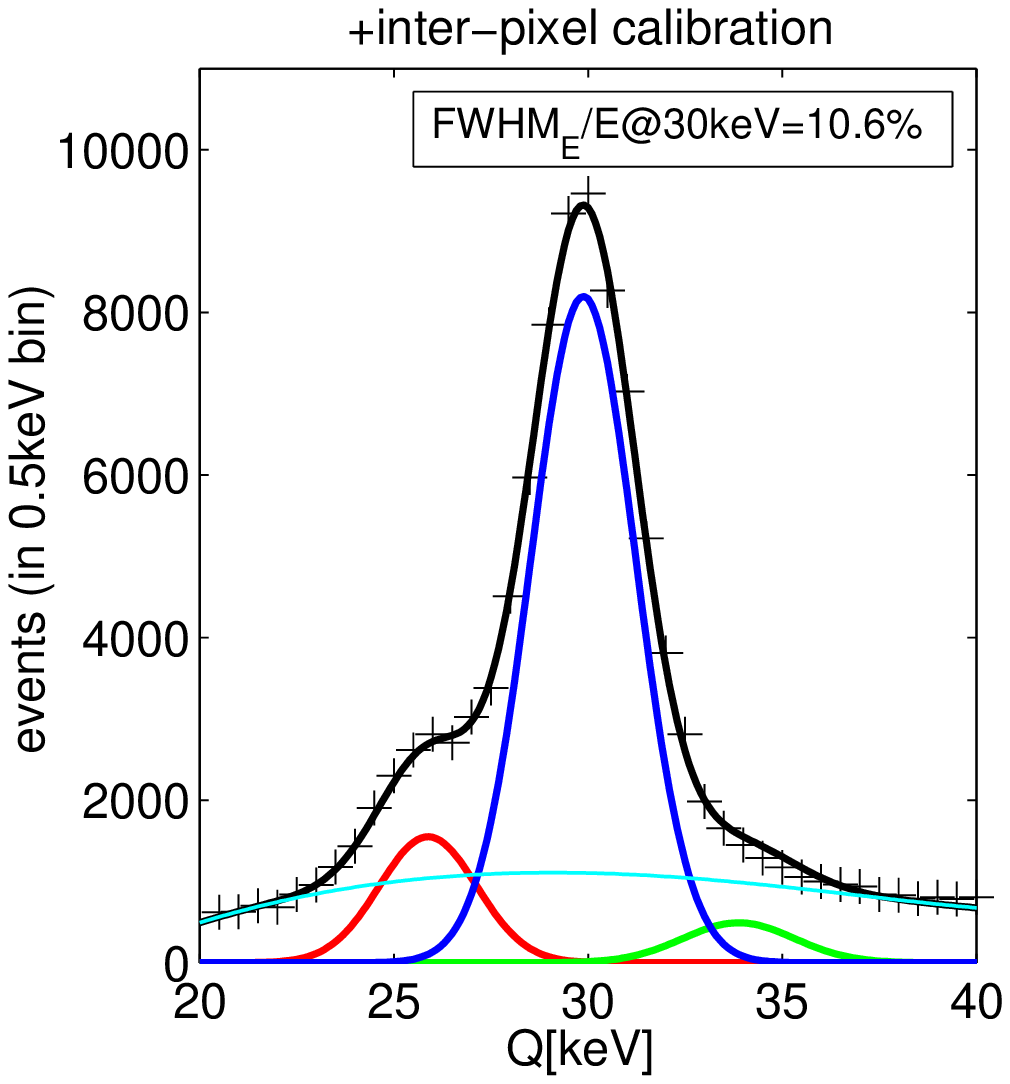}
\includegraphics*[width=4.5cm]{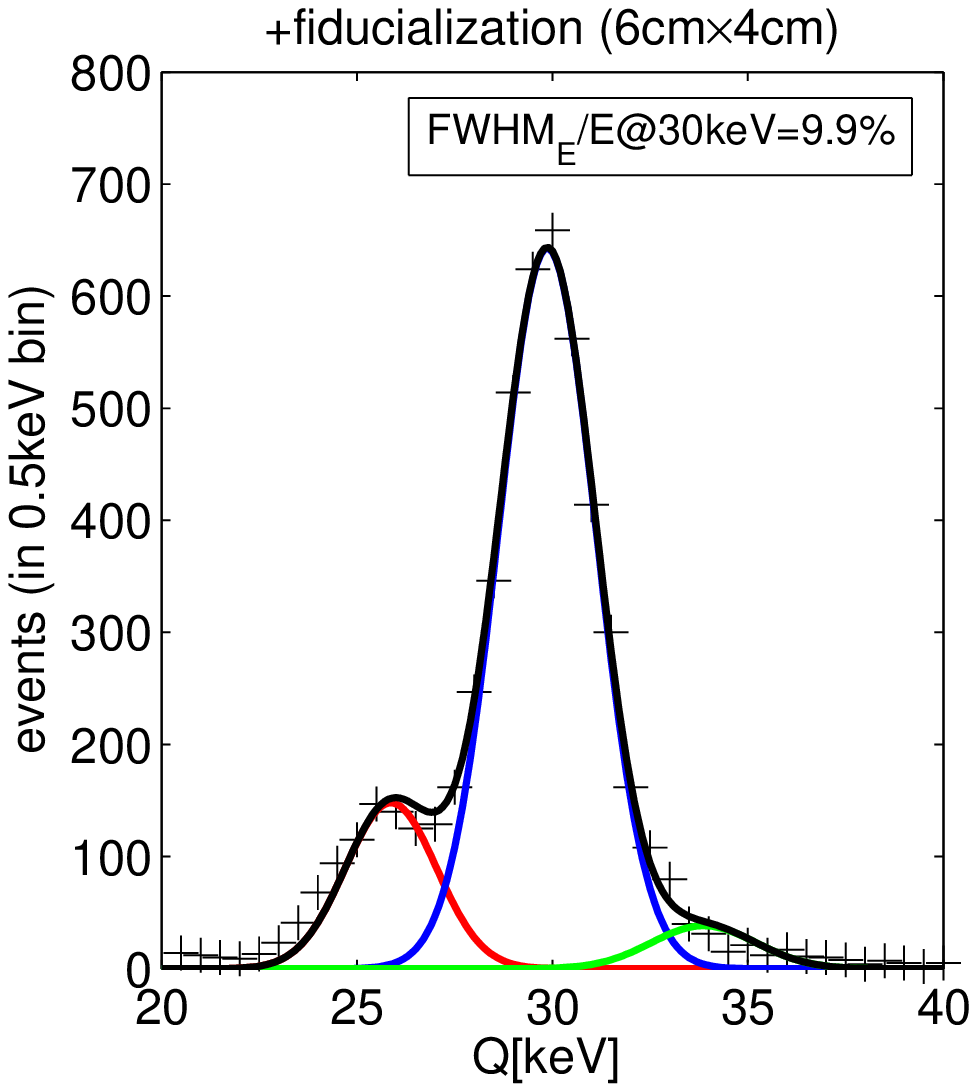}

\includegraphics*[width=4.5cm]{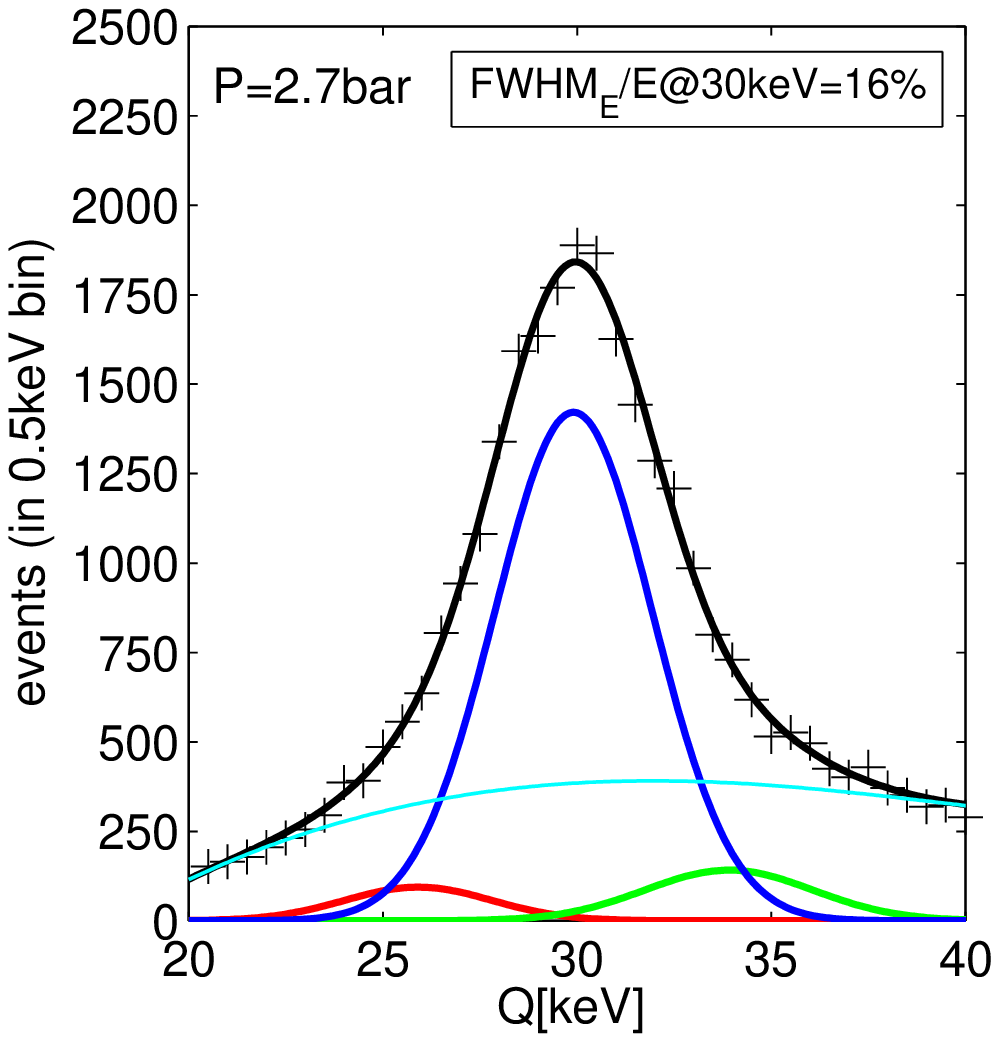}
\includegraphics*[width=4.5cm]{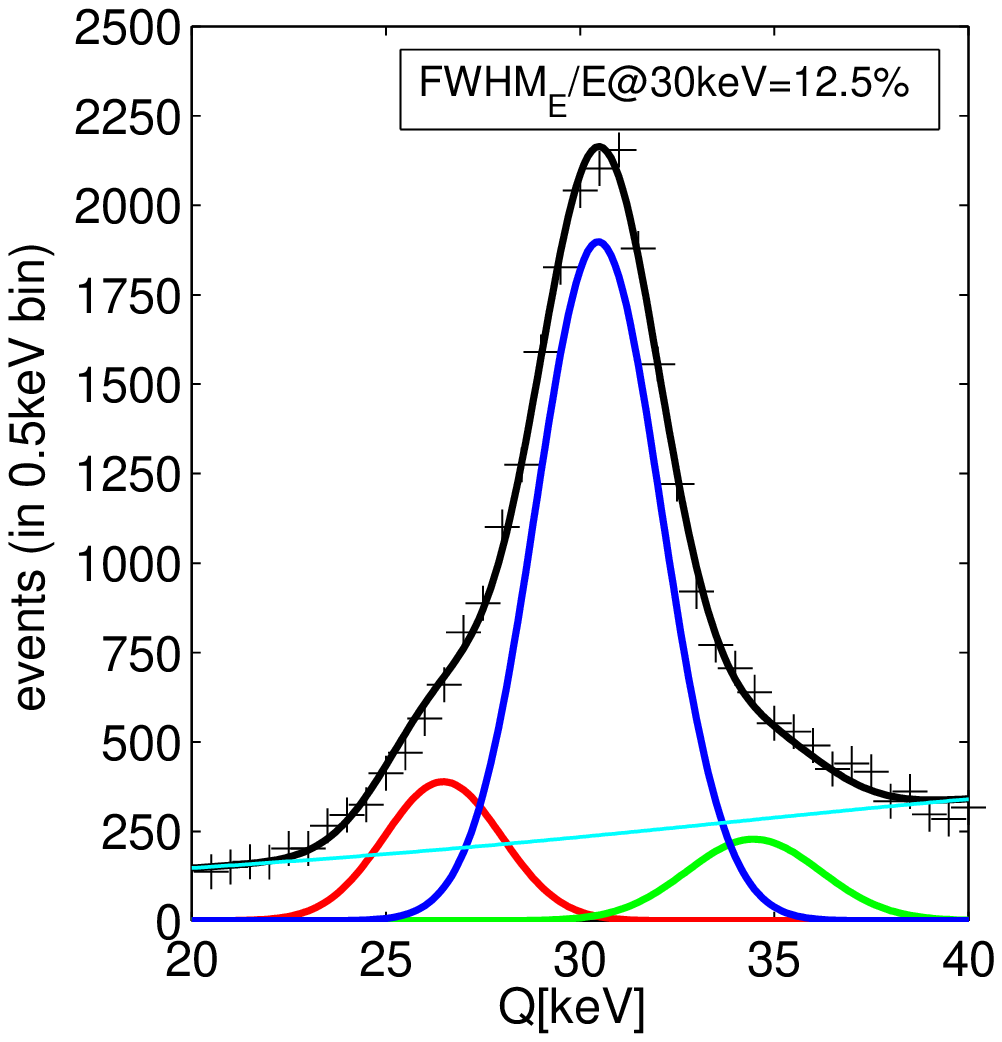}
\includegraphics*[width=4.5cm]{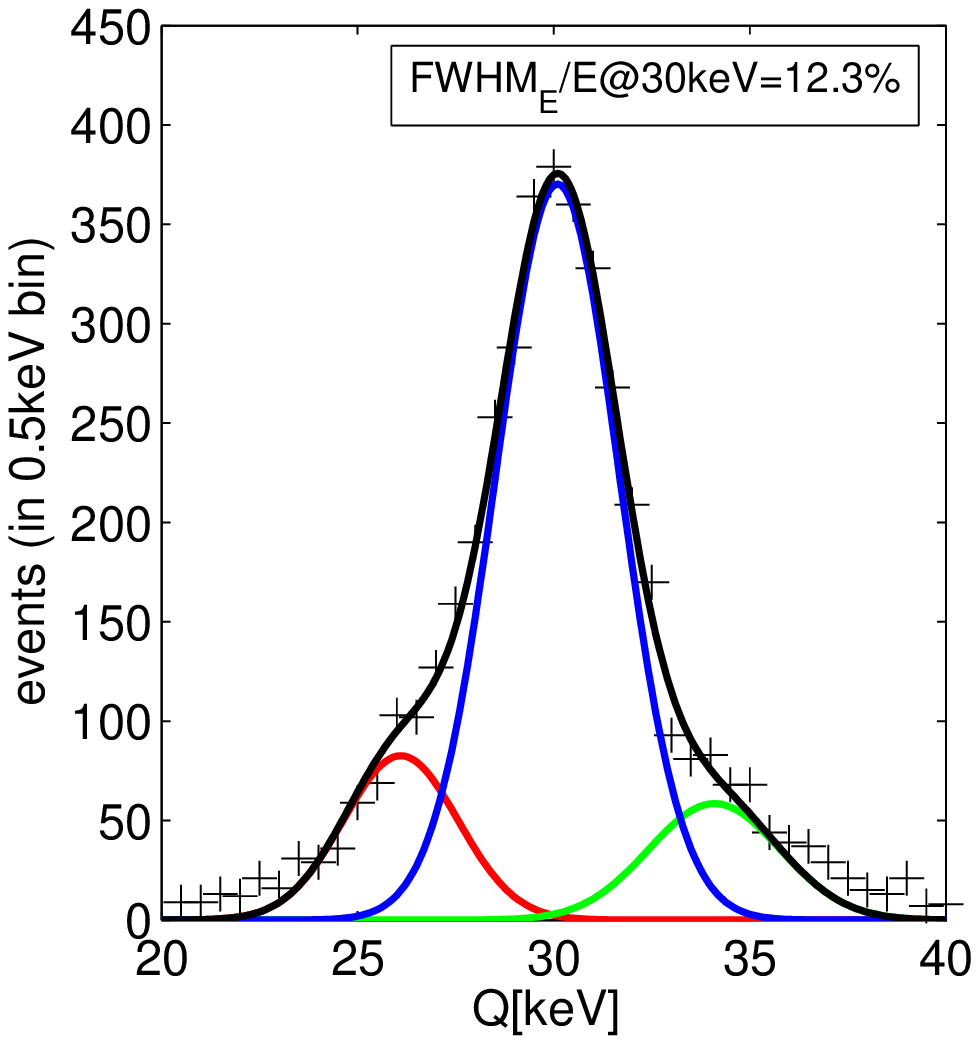}
\caption{From left to right: energy resolutions after calibrating the different quadrant gains, after calibrating the different pixel gains, after selecting a fiducial region, applied in a cumulative way. Results for P=1\,bar (upper row) are obtained at $E_{MM}/P=54$\,V/cm/bar and $E_{drift}/P=145.5$\,V/cm/bar (2.2\% of TMA). Results for P=2.7\,bar (lower row) are obtained at $E_{MM}/P=25$\,V/cm/bar and $E_{drift}/P=104$\,V/cm/bar (2.4\% of TMA). The red Gaussian function describes the secondary $\gamma$-emission from $^{237}$Np super-imposed on the $K_{\beta}$ escape peak, and the green Gaussian function represents the corresponding `orphan' $K_{\beta}$ X-ray arising from primary interactions outside the chamber active volume. The main $K_{\alpha}$ escape peak super-imposed on the direct $K_{\alpha}$ emission is shown in dark blue. The background (cyan) is described by a $3^{rd}$ order polynomial.}
\label{Cal_plots}
\end{figure}

The energy resolution is an important figure for assessing the system behavior. A different response can be a priori expected for each quadrant, for different regions within a quadrant, or for positions close to the boundaries of the chamber (or quadrants). Its understanding is important for the characterization of the properties of the mixture itself that is attempted in the next section. To that purpose, an energy-calibration algorithm has been developed, consisting of the following steps:
\begin{enumerate}
\item Inter-quadrant calibration: gain variations from quadrant to quadrant were corrected in a first step by using global factors applied to events contained in each of them. Charge from events shared between quadrants was not corrected. The total spectrum was then calibrated assuming the position of the low-energy peak to be placed at 30\,keV. The result from this step is shown in Fig. \ref{Cal_plots}-left. It must be noted that the (verified) accuracy of the HV supply used for the readout is around $\pm 2$\,V (translating into 1\% for these measurements). It is unclear at the moment whether the observed 10\% gain variations between quadrants are related to the HV supply or they originated during the manufacturing process of the Micromegas.
\item\label{cal1} Inter-pixel calibration: a matrix of charge spectra was created, with one spectrum per pixel. Each spectrum was filled whenever the given pixel collected the highest charge of the event.
\item\label{cal2} Gaussian fits were performed to the spectra in the 30\,keV region. The mean value obtained from the fit was normalized to the mean value obtained over the full plane, returning a gain map. Only spectra containing more than 25\,evts were fit, for pixels having between 15 and 25\,evts a simple mean was used, and below 15\,evts no action was taken. This procedure allowed the calibration of more than 95\% of the instrumented readout plane for the cases here studied, with a relative gain spread amounting to some 20\% typically.
\item After producing the gain map all pixel charges can be corrected and the procedure iterated starting from \ref{cal1}. For simplicity, the calibration of the 30\,keV peak was performed in this work without any iteration and the gain map obtained after \ref{cal2} introduced as a single correction factor for the event.
\end{enumerate}

Results are shown in Fig. \ref{Cal_plots} after the inter-quadrant calibration (left), inter-pixel calibration (middle), and $x$-$y$ fiducialization in a 6\,cm$\times$4\,cm region, applied cumulatively. Background around $30\,$keV stems presumably from partially-contained 59.5\,keV photons, since it almost completely disappears after $x$-$y$ fiducialization. The spectra are fitted to 3 Gaussian peaks (red, blue, green) with energy differences given by table \ref{sources} and energy resolutions bound through $1/\sqrt{E}$ scaling. In this way only the amplitudes of each Gaussian peak as well as the position and width of the 30\,keV peak are floating during the fit, together with a $3^{rd}$ order polynomial used for describing the background (cyan).

The aim of the calibration procedure at this stage was mainly to ensure a correct interpretation of the spectrum and to help suppressing backgrounds before proceeding with the extraction of the parameters of the electron swarm for the Xe/TMA mixture, that is attempted in next section. The obtained energy resolution is indeed within a factor 1.7 of the best ones obtained in small amplifying structures of this type, reported in \cite{Diana1}. We foresee an improvement in the calibration algorithms avoiding some of the shortcuts here followed, including pulse shape analysis at lower thresholds as well as optimizing the Micromegas working point (transparency and gain) in order to approach the resolutions obtained in small setups. On another hand, it is expected from \cite{NEXT_DEMO} that the proposed calibration algorithm can be generically applied in Xenon mixtures to any arbitrary ionization trail and pressure, provided the characteristic $K_\alpha$ emission can be isolated. The remarkable fact that the main escape peak in Xenon for $^{241}$Am $\gamma$'s coincides with its $K_\alpha$ line makes this source extremely suited and valuable for further testing the calibration algorithms at low pressures.
The correct energy scale needed for calibration emerges naturally and no additional pattern recognition is needed to identify the displaced ionization, contrary to genuine $K_\alpha$ emission in association with the primary track.



\section{Extraction of the parameters of the electron swarm}\label{swarm}
\begin{figure}[h]
\centering
\includegraphics*[width=6.5cm]{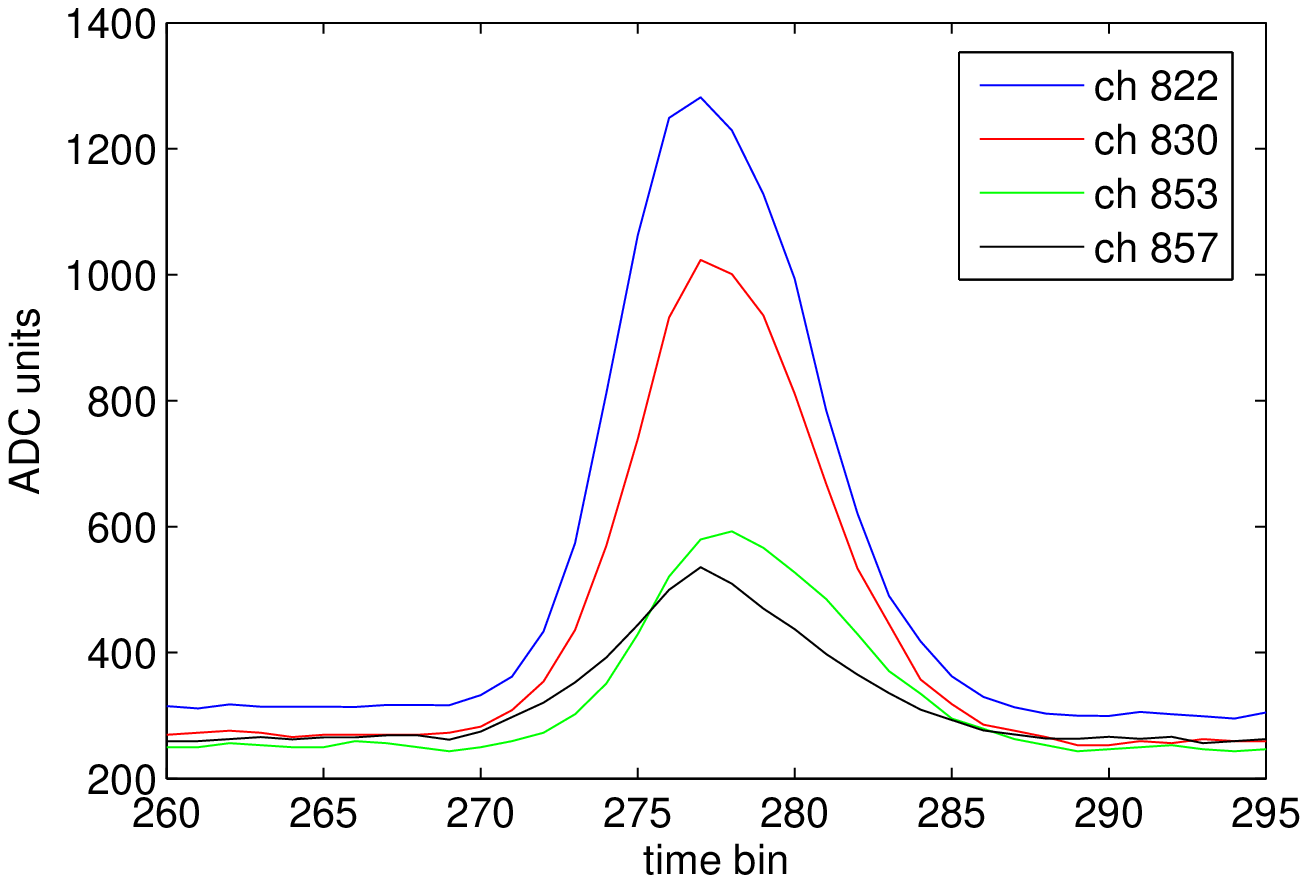}
\includegraphics*[width=6.5cm]{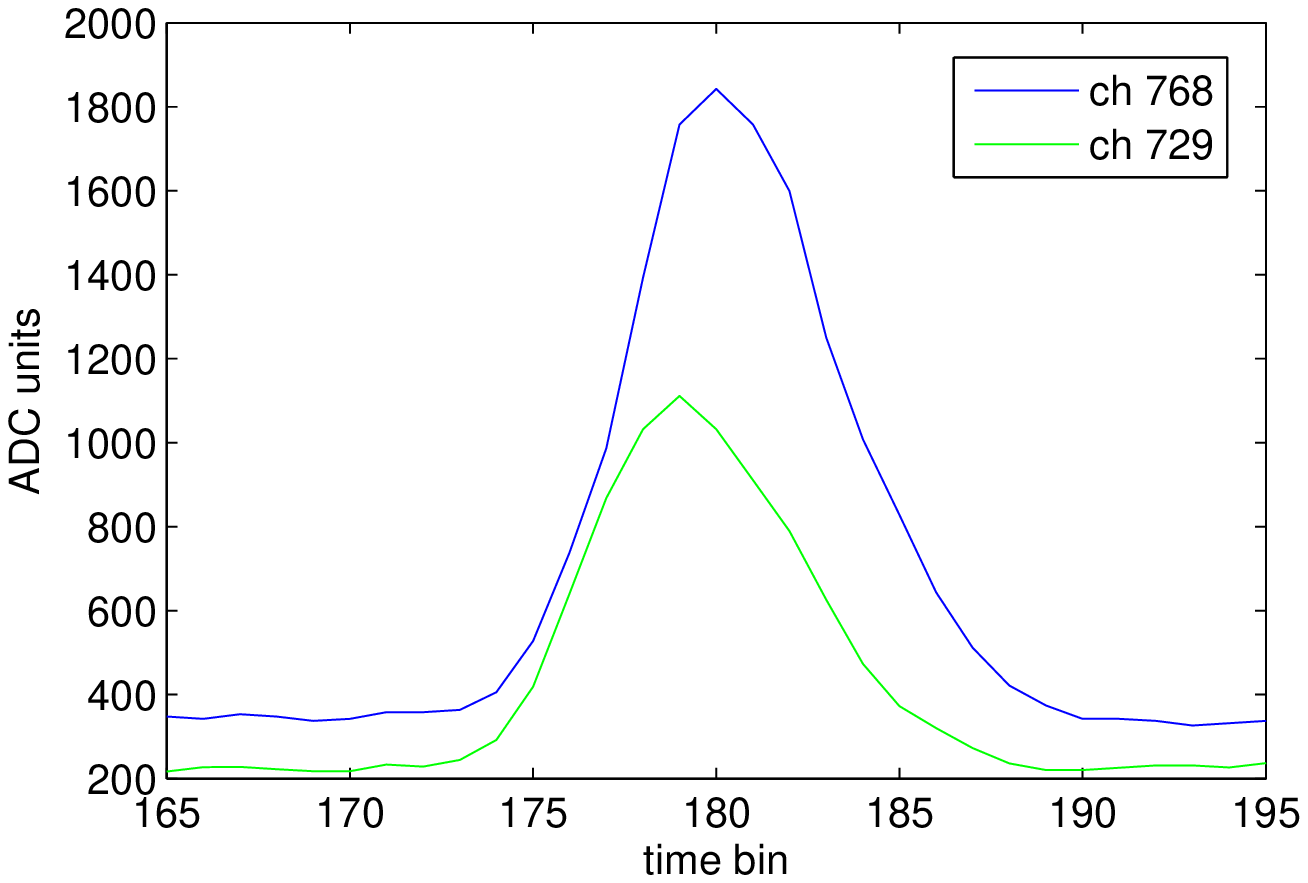}
\caption{Events obtained from a high purity $\gamma$-ray sample in a narrow energy window $25\,$keV$<\epsilon_{evt}<35\,$keV after $x$-$y$ and $z$-fiducialization, and after performing a $width-t\!ime$ correlation cut (see the text for details).}
\label{PSA_all}
\end{figure}
\subsection{Pulse-shape analysis}
Two typical events recorded around the 30\,keV escape peak energy are shown in Fig. \ref{PSA_all}, sampled with a time bin of $0.5\,\mu$s. In the default pulse-shape analysis (PSA) signals recorded for each channel/pixel above a certain threshold are fitted to a sum of Gaussian functions:
\beq
I_{i,fired}(t) = I_{i,0} + \sum_{j=1}^{N_{Gaussians}} A_{\!j} ~ e^{-\frac{(t-\bar{t}_{\!j})^2}{2\sigma_{t,j}^2}} \label{I_gauss}
\eeq
where the channel pedestal, $I_{i,0}$, is determined before hand on an event-by-event and channel-by-channel basis. This multi-Gaussian analysis increases the stability of the fitting procedure: it can capture easily double-clusters from 59.5\,keV $\gamma$'s that are originated at about the same pixel but different $z$-coordinate, for instance.

A threshold corresponding to approximately 1\,keV (100 ADC units), relative to the pedestal estimate, was used. In order to avoid any residual ballistic deficit when estimating the event charge from the pulse amplitude, the former was re-obtained as the analytical integral of all fitted Gaussians above threshold. The charge was translated into the energy of the event, $\epsilon_{evt}$, by using the 30\,keV peak as a reference. Positions in the $x$-$y$ plane were obtained from the event barycenter following eqs. \ref{x_evt}, \ref{y_evt}, \ref{r_evt}.

Naturally, the values of $\bar{t}_{j}$ and $\sigma_{t,j}$ obtained from the fit to eq. \ref{I_gauss} are related to the electron drift velocity and longitudinal diffusion. In order to minimize the contribution of the size of the initial ionization cloud to the estimate of these parameters we focus on 30\,keV events, in particular on the pixel carrying the highest fraction of the event charge (about 70\% in average, Fig. \ref{Dl_Dt}-right). Signals from those pixels are nearly always described by a single Gaussian fit and hence they allow for the simplified notation $t_{M}\equiv \bar{t}_j$ and $\sigma_{t,M}\equiv \sigma_{t,j}$, used hereafer.

\begin{figure}[h]
\centering
\includegraphics*[width=7.5cm]{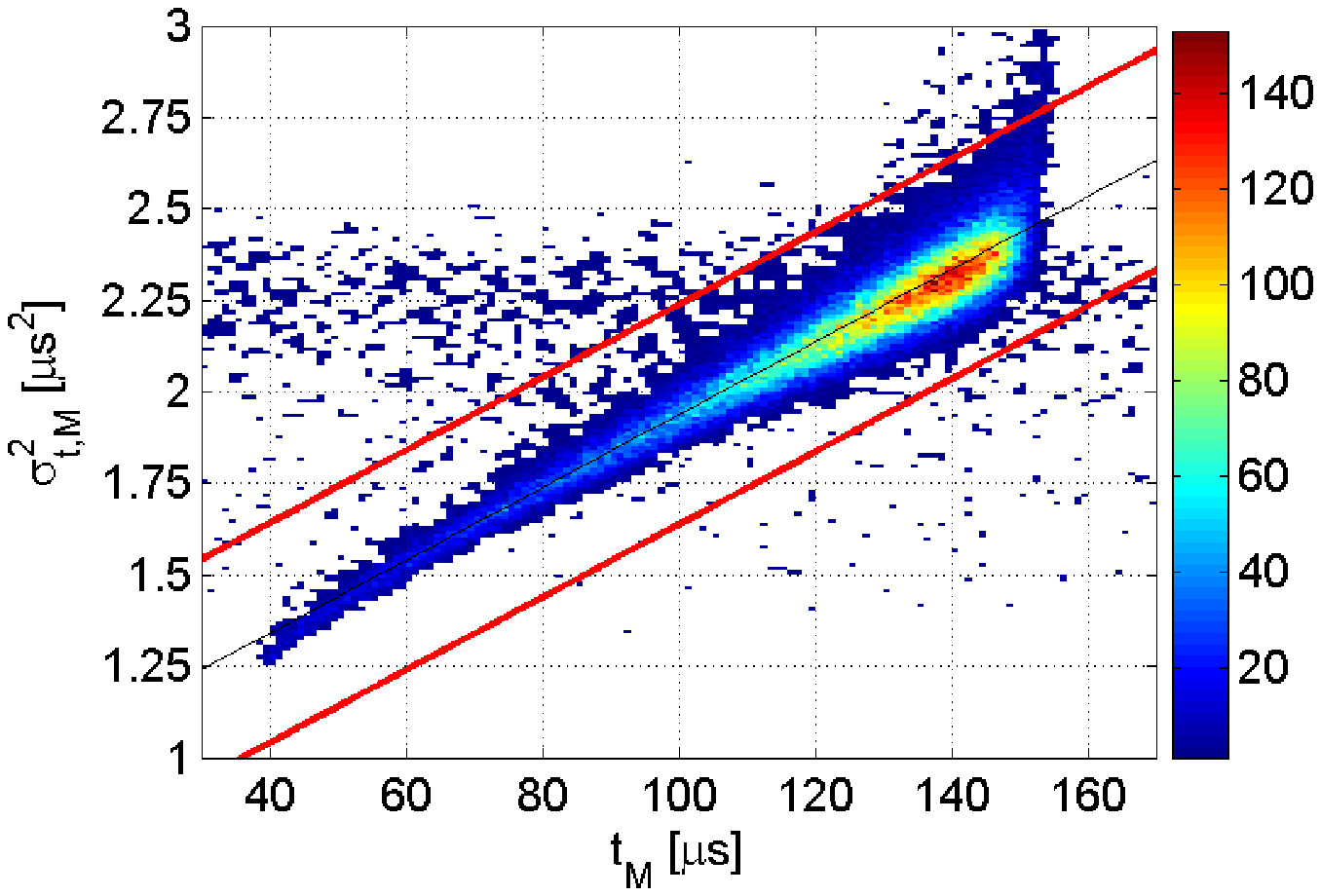}
\includegraphics*[width=7.5cm]{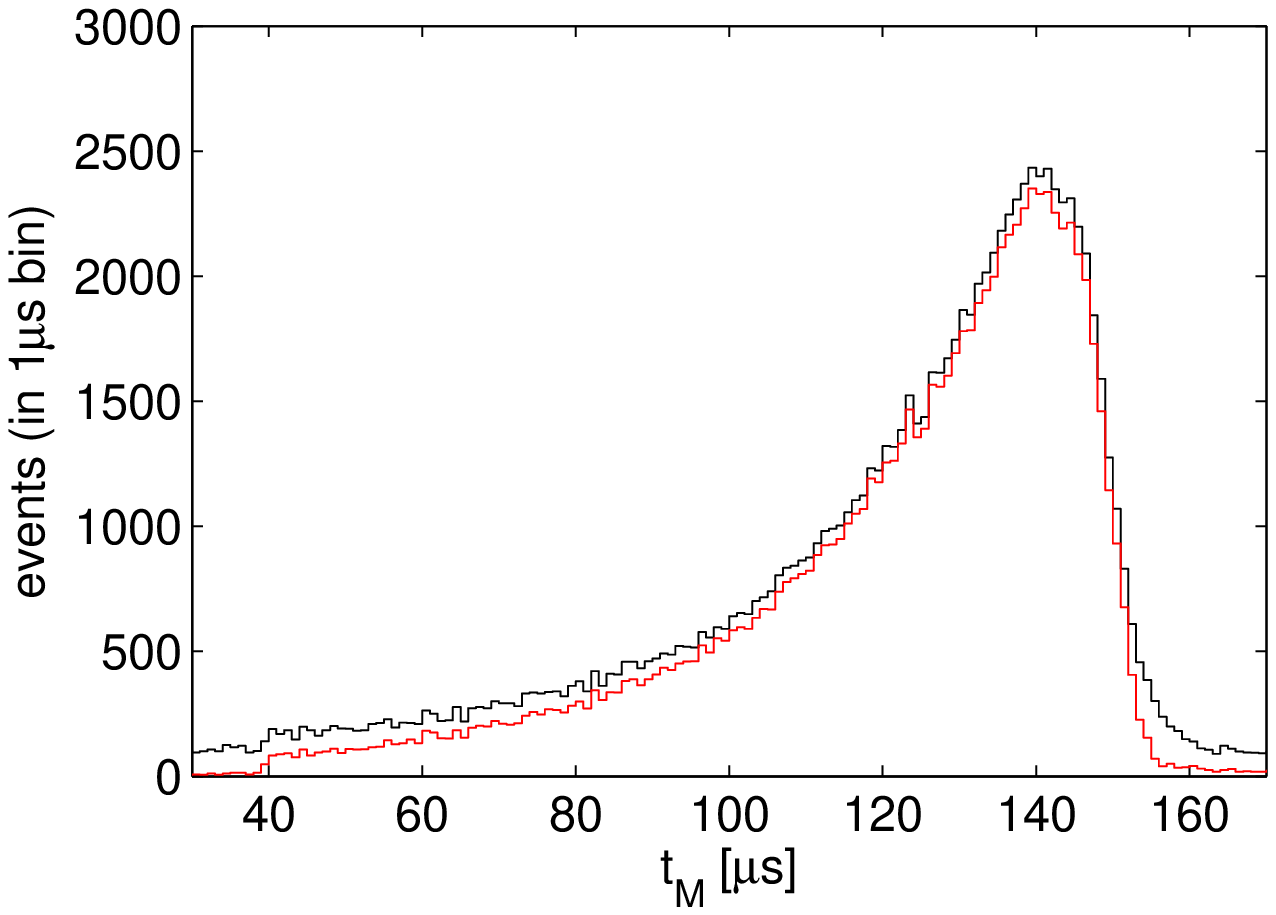}
\caption{Timing characteristics of localized energy deposits in the TPC at $E_{drift}=145.5$\,V/cm ($P=1\,$bar) for a mixture of Xe/TMA in proportion 97.8/2.2. Left: width-time correlation curve ($\sigma^2_{t,M}-t_{M}$) for the pixel containing the highest fraction of the event charge, for events gated around the escape peak energy ($25\,$keV$<\epsilon_{evt}<35\,$keV). Right: time distribution for those events before (black) and after (red) the correlation cut (red lines on left-figure) used to suppress random coincidences from the source itself. The magnitude $t_{M}$ has an offset relative to the start-time of the DAQ window. The offset in $\sigma^2_{t,M}$ is mainly due to the electronics response function, with a small contribution coming from the size of the ionization cloud.}
\label{W_t_corr}
\end{figure}

The $\sigma^2_{t,M}-t_{M}$ characteristics of 30\,keV events in the TPC are shown for illustration in Fig. \ref{W_t_corr}-left, for a drift field $E_{drift}=145.5$\,V/cm and a mixture of Xe/TMA at 97.8/2.2 and $P=1$\,bar. The observed straight line is expected from longitudinal diffusion:
\beq
\sigma^2_{t,M}=\sigma^2_0 + \frac{2D_L}{v_d^2}t_M \label{diffL}
\eeq
with $D_L$ referring to the longitudinal diffusion coefficient and $v_d$ to the electron drift velocity. The offset $\sigma^2_0$ contains the effect of the convolution with the electronics response function and, to a smaller extent, the space extension of the ionization cloud, both being $z$-independent (see the discussion section). The $\sigma^2_{t,M}-t_{M}$ correlation is so strong in present conditions that all remaining random coincidences with events originated at the source can be easily suppressed (Fig. \ref{W_t_corr}-right).


\subsection{Determination of the total Drift Time}

\begin{figure}[h]
\centering
\includegraphics*[width=7.7cm]{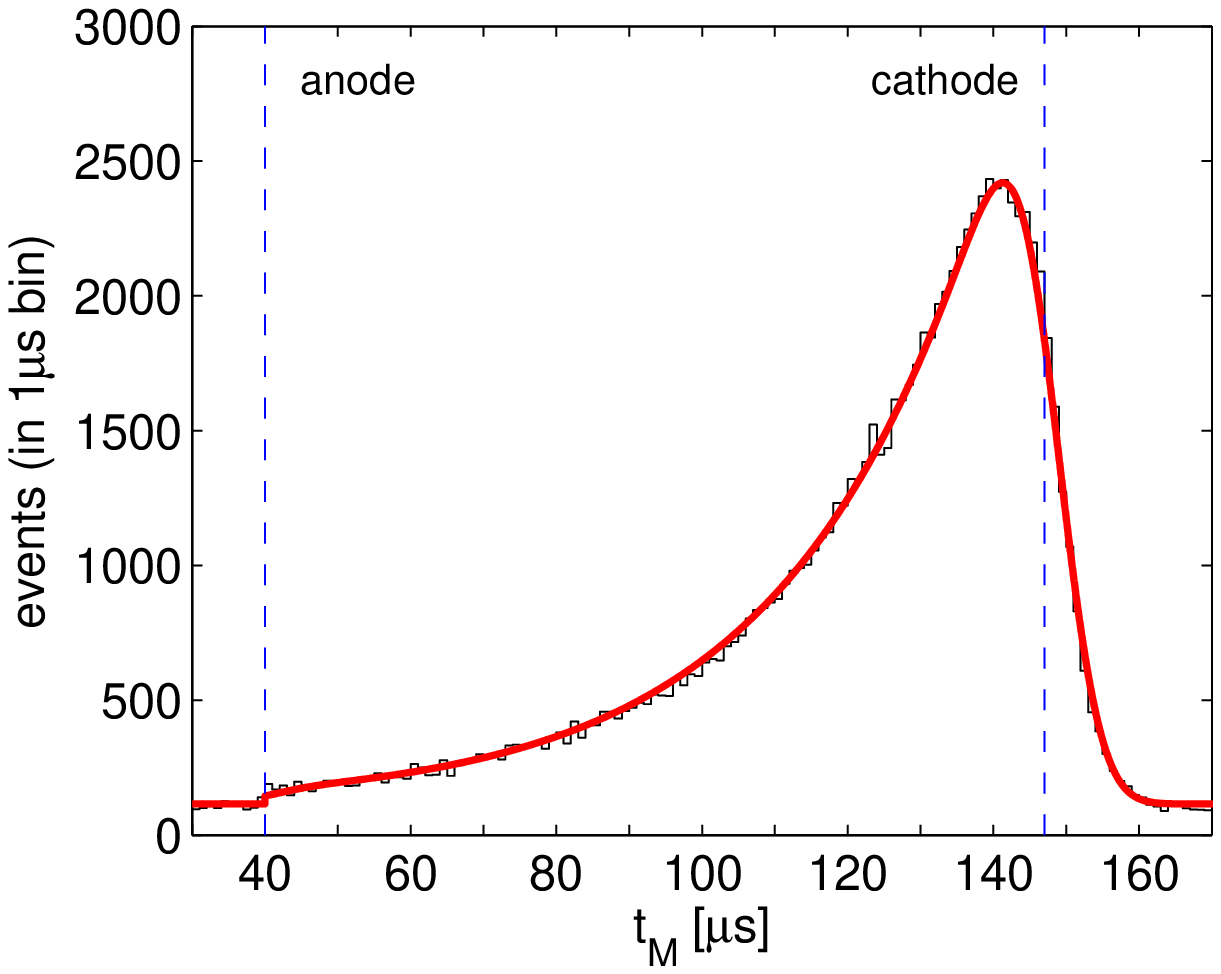}
\includegraphics*[width=6.7cm]{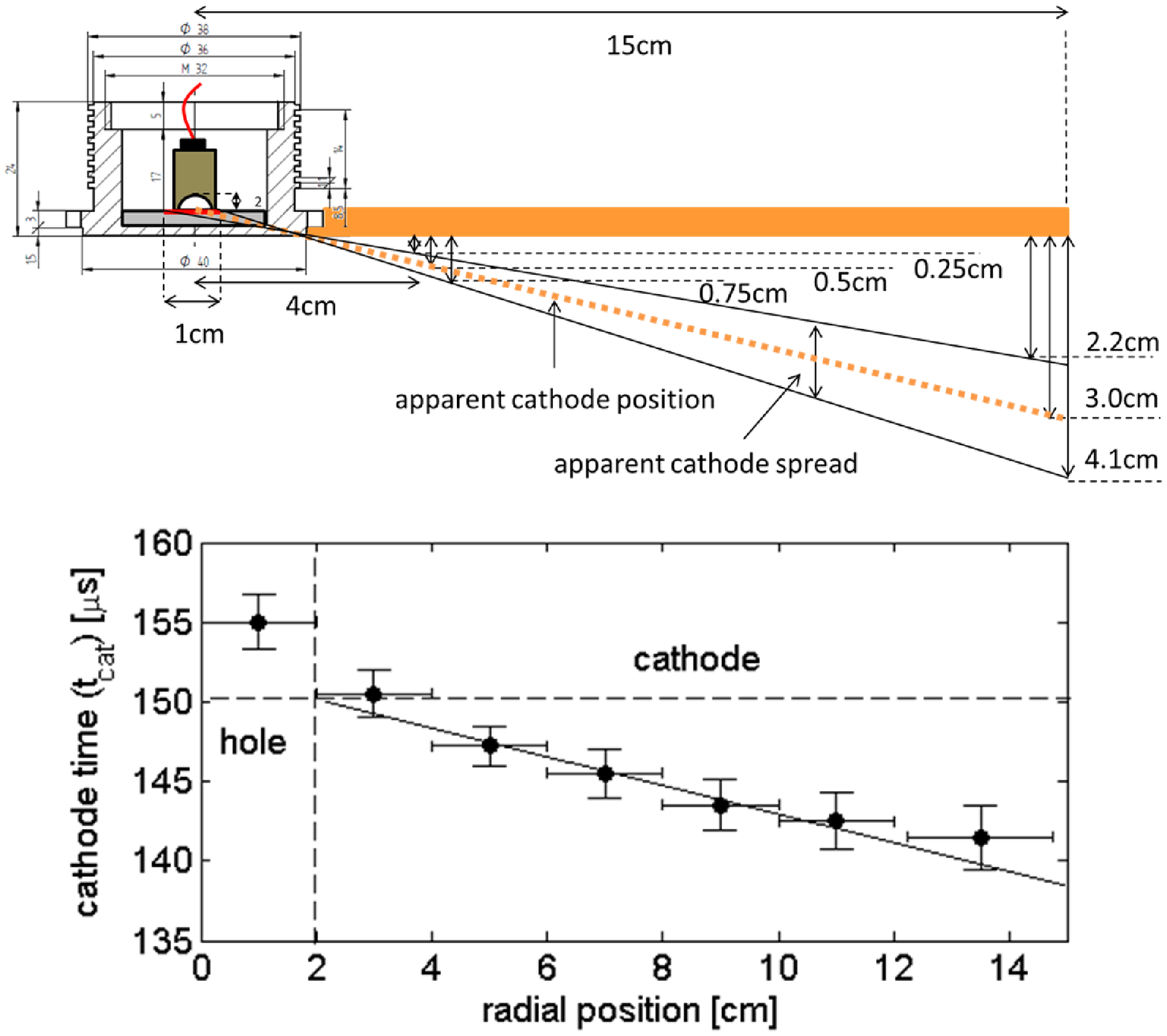}
\caption{Sketch of the procedure used for determining the TPC drift region under $\gamma$ irradiation at $E_{drift}/P=145.5$ V/cm/bar, $P=1$\,bar. Left: time distribution and fit to the function introduced in the text (red line). Right-up: approximate sketch of the geometry in the source region. Right-down: the apparent position of the cathode for different slices in the radial position of the energy deposit ($r_{evt}$) as obtained from the left fit. The estimate of the cathode position ($t_{cat}=150\,\mu$s) from this differential analysis differs by about 2\% with respect to a global fit ($t_{cat}=147\,\mu$s).}
\label{Drift}
\end{figure}

The time ($t_M$) distributions obtained for 30\,keV events are characterized by a sharp left edge, attributed to the readout plane, and by a relatively smooth right edge near the cathode. The cathode edge is smoothed due to the `shadow' created by the extended $^{241}$Am source (1\,cm diameter) when illuminating the cathode hole edge. Additionally, for radii smaller than the cathode hole ($r_{evt}<2$\,cm) fringe fields slow down the primary electron swarm, further enhancing the time spread (Fig. \ref{Drift}-right). For the extraction of the parameters of the swarm as here intended, a fiducial cut was applied by removing events having any charge deposit within 2.5\,cm of the boundaries of the active region of the readout plane (given by Fig. \ref{examples_P}-left), leaving a usable area of around 400\,cm$^2$ for the present analysis.

A convenient parameterization of the time distribution can be found by resorting to the convolution of an exponential function (approximating the physical distribution of events along the drift region) with a Gaussian distribution that characterizes the aforementioned geometrical effect through its width $\sigma_g$. The resulting distribution can be expressed as the difference of two error functions (erf):
\beq
f(t_M) = \mathcal{C} e^{t_M/\tau^*} \left[\tn{erf}\left( \frac{t_{cat}-\sigma_g^2/\tau^*-t_M}{\sqrt{2}\sigma_g}\right)- \tn{erf}\left( \frac{t_{ano}-\sigma_g^2/\tau^*-t_M}{\sqrt{2}\sigma_g}\right)\right]\Theta(t_M-t_{ano}) + B \label{z_fit}
\eeq
The pre-factor $\mathcal{C}$ stands for a normalization constant, $B$ is the background level and $t_{cat(ano)}$ refer to the absolute times (provided by the DAQ) assigned to the cathode and anode edge, respectively. The dominant attenuation law behavior is parameterized through an effective electron life-time $\tau^*$. A sharp step-function $\Theta(t_M-t_{ano})$ is introduced at the anode edge as seen in data.

An example of a typical fit to eq. \ref{z_fit} is shown in Fig.\ref{Drift}-left for a drift field $E_{drift}=145.5$ V/cm and 1\,bar pressure, yielding a drift region enlarged by about 5\% with respect to a naive estimate from the position of the cathode peak.
Besides smoothing the cathode region, the geometrical `shadow effect' causes an apparent shift in the cathode position as a function of the radial coordinate of the charge deposit, approximately following a straight line (Fig. \ref{Drift}-right). This makes the estimate of the drift region differ by about 2\% relative to that obtained through a direct fit performed without any radial selection, and so  $t_{cat}$ must be corrected accordingly. The time span of the drift region can be finally obtained as $\Delta T = t_{cat}-t_{ano}$.

\subsection{Drift velocity, diffusion and attachment coefficients of Xe/TMA mixtures}

The statistical uncertainty on the determination of the drift region resulting from the fit to eq. \ref{z_fit} as well as the mechanical accuracy in the drift region ($D=38$\,cm) are around 1\%, that directly translate into the corresponding uncertainty in the drift velocity $v_d = D/\Delta T$. Nevertheless, the present uncertainty of our experimental setup is presumably dominated by the procedure used to identify the cathode position, therefore we assign in the following a maximum 5\% systematic uncertainty to $v_d$, the typical difference observed between an estimate from the fit and one from the cathode peak position. More systematic studies and/or a single measurement in a well known gas will expectedly reduce the present uncertainty in future measurements to near the statistical levels.

\begin{figure}[h]
 \centering
 \includegraphics*[width=7.5cm]{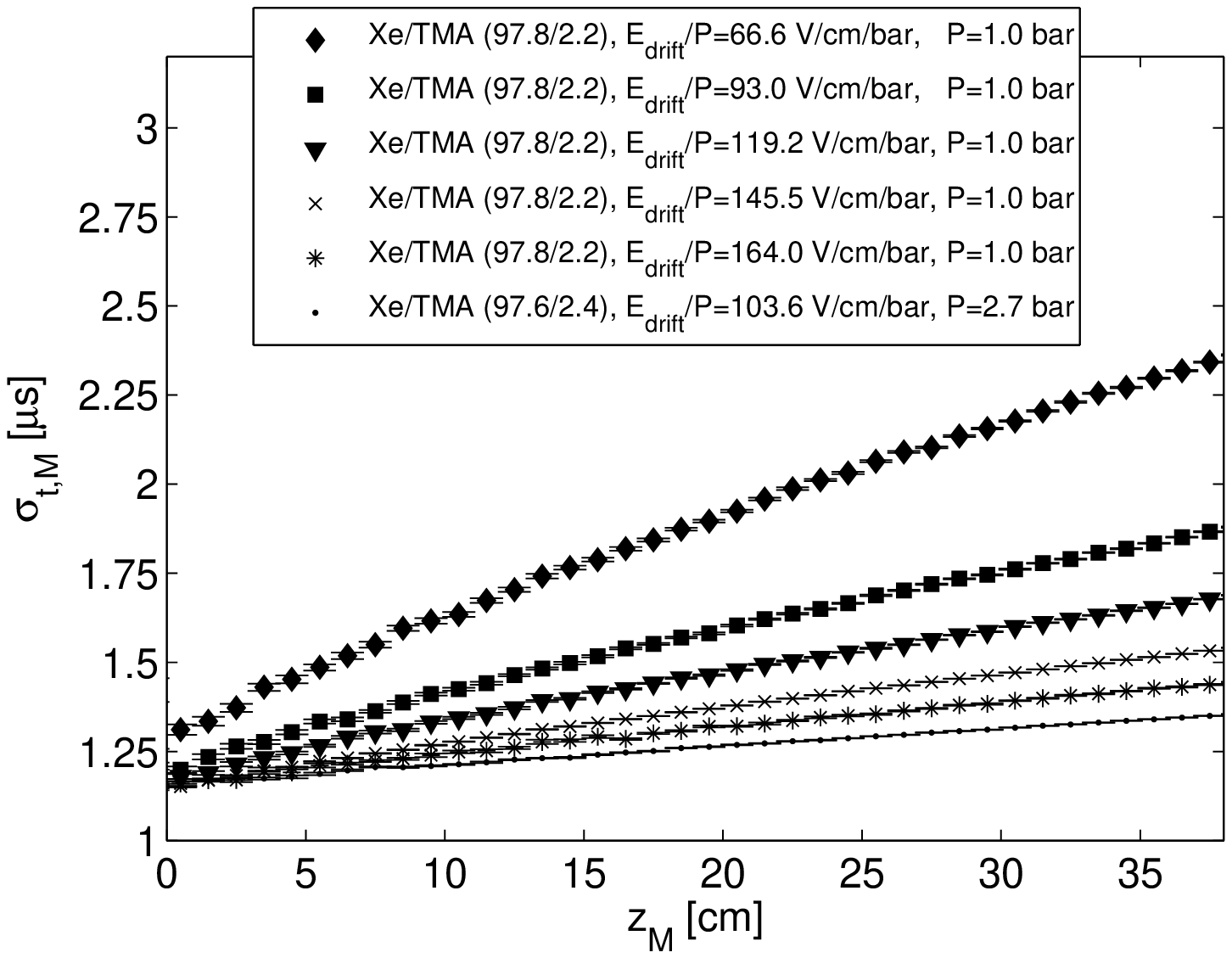}
 \includegraphics*[width=7.5cm]{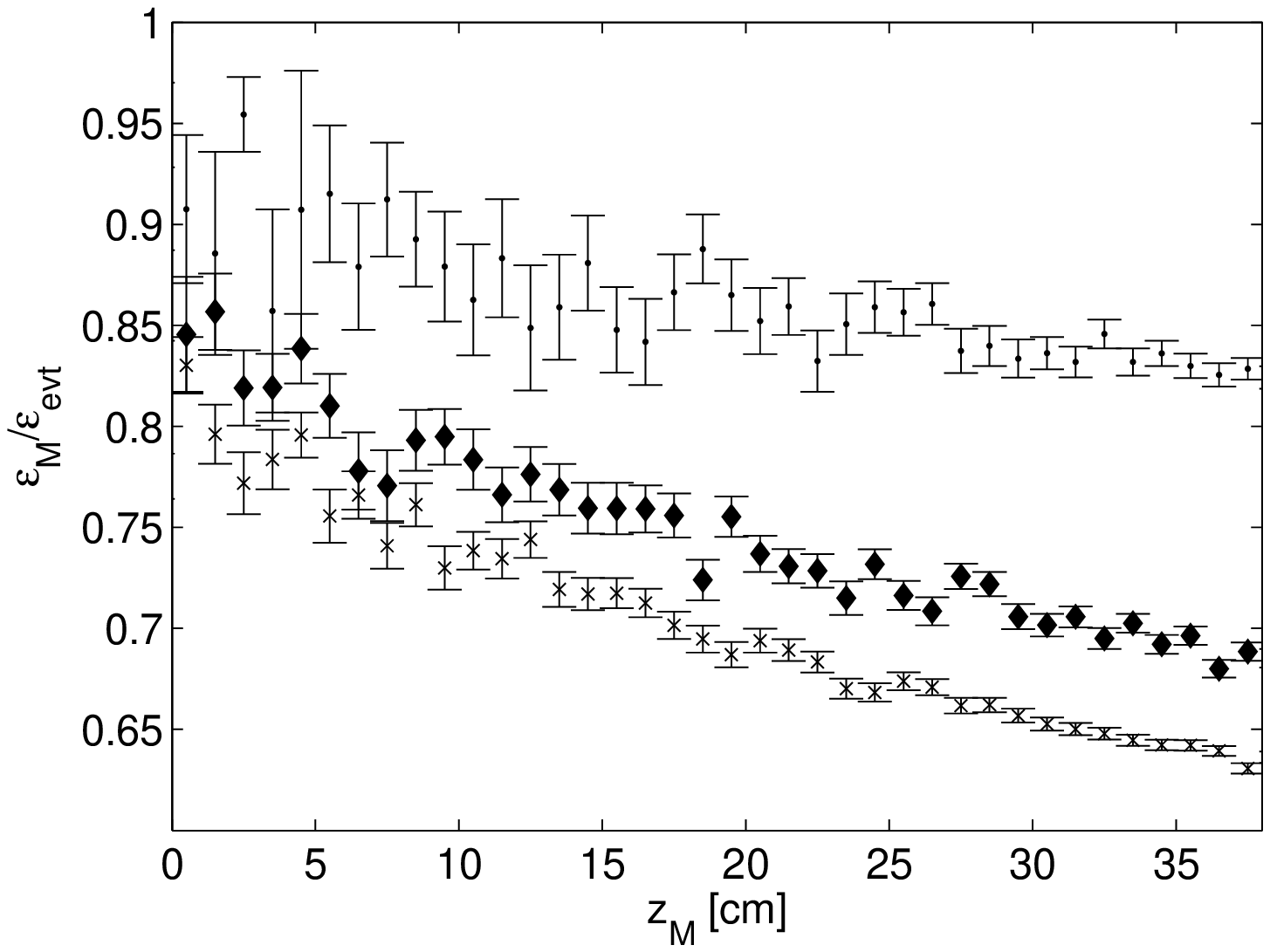}
 \caption{Left: width-time correlation in $z$-slices for all runs used in this section. Right: highest charge fraction contained in a single pixel for a 30\,keV event as a function of the position of the charge deposit within the drift region. The decreasing trend towards high $z$ values indicates the presence of transverse diffusion: the highest reduction seen for about the highest field, the lowest reduction seen for the highest pressure (nearly flat), following the behaviour shown in left figure. For increased clarity only 3 data sets are shown.}
 \label{Dl_Dt}
 \end{figure}

By studying the chamber behaviour as a function of the drift distance, $z_M=v_d\times (t_M-t_{ano})$, it is possible to extract information about the diffusion and attachment coefficients. Fig. \ref{Dl_Dt}-left shows the behaviour of $\sigma_{t,M}-z_{M}$ in narrow slices obtained for various runs that are later used for the extraction of the longitudinal diffusion coefficient. The reduction of diffusion with increased pressure (points) is apparent, as expected. Another relevant feature is that for most of the data sets, except for $E_{drift}/P=$66.6 V/cm/bar (diamonds), a consistent value of the vertical intercept ($\sigma_{0}$) is observed (the deviation arises from the much reduced drift velocity, a fact that has been used in the discussion section to estimate the size of the initial ionization cloud). The previous observation suggests that $\sigma_{0}$ stems fundamentally from the electronics response function with little contribution from the ionization cloud (whose time duration is both field and pressure dependent). This hypothesis was partly addressed for a reference case ($E_{drift}=145.5$\,V/cm/bar) by using for the PSA a computing-intense fit to all pulses based on a numerical convolution of a Gaussian function with the known FEE response function (obtained from \cite{Baron}), characterized by a shaping time $\tau$. The $\sigma_{t,M}-z_{M}$ slope obtained in this way agreed with the one coming from a standard Gaussian PSA (eq. \ref{I_gauss}) within 5\%.

The fit of the $\sigma_{t,M}-t_{M}$ correlation plots (as those of Fig. \ref{W_t_corr}) to eq. \ref{diffL} returns a slope given by $\sqrt{\frac{2D_L}{v_d^2}}$ without previous knowledge of the drift velocity $v_d$. This is to be preferred to a $\sigma_{t,M}-z_{M}$ fit since it simplifies the propagation of uncertainties. Hence, starting from the known value of the velocity $v_d$ and this slope, the longitudinal diffusion coefficient can be defined for convenience as:
\beq
D^*_{L}=\sqrt{\frac{T_0}{T}\frac{2P ~ D_{L}}{v_d}} ~~~~[\frac{\mu\tn{m}}{\sqrt{\tn{cm}}}\times\sqrt{\tn{bar}}] \label{Deff}
\eeq
with T$_0=20\,^\circ$C and $D_{L}$ evaluated at this reference temperature. An analogous definition is followed for the transverse diffusion coefficient $D^*_{T}$ since both coefficients become in this way independent from the gas pressure, $P$, and allow for a simpler comparison with existing data.\footnote{In case of non-ideal gas behavior, the gas density $N$ is a more suited variable. Since the compressibility factor of pure Xenon nears 0.94 even at 10\,bar (0.995 at 1\,bar) and its exact value for Xe/TMA is unknown, we avoid this complexity in the following.} The spread of a point-like ionization cloud propagating along a distance $z$ can be obtained after the definition in eq. \ref{Deff} as:
\beq
\sigma_{L,T} = D^*_{L,T} \frac{\sqrt{z}}{\sqrt{P}} \label{diff_vs_P}
\eeq
The transverse diffusion coefficient $D^*_T$ can be a priori studied thanks to the segmentation of the readout plane, however the modest pixel multiplicity of 2-4 (Fig. \ref{examples_P_1D}) in the present case complicates the standard analysis based on a barycenter calculation. In order to illustrate the trend we chose to study the behaviour, as a function of the drift distance,
of the highest charge fraction per pixel $\epsilon_M/\epsilon_{evt}$ (Fig. \ref{Dl_Dt}-right). A high purity sample was provided by applying a $width$-$time$ correlation cut (Fig.\ref{W_t_corr}-left) to remove the biasing contribution of random coincidences. The average value of $\epsilon_M/\epsilon_{evt}$ is indeed reduced as a function of the distance to the readout plane as expected from the increased transverse diffusion. This observable, being the ratio of two charges, is immune to attachment and is not sensitive to longitudinal diffusion. The trends show how the transverse size of the ionization cloud decreases with pressure, as given by eq. \ref{diff_vs_P}, and increases with the drift field similarly to the behaviour of the longitudinal diffusion in this range of reduced fields and TMA admixtures.

\begin{figure}[h]
\centering
\includegraphics*[width=\linewidth]{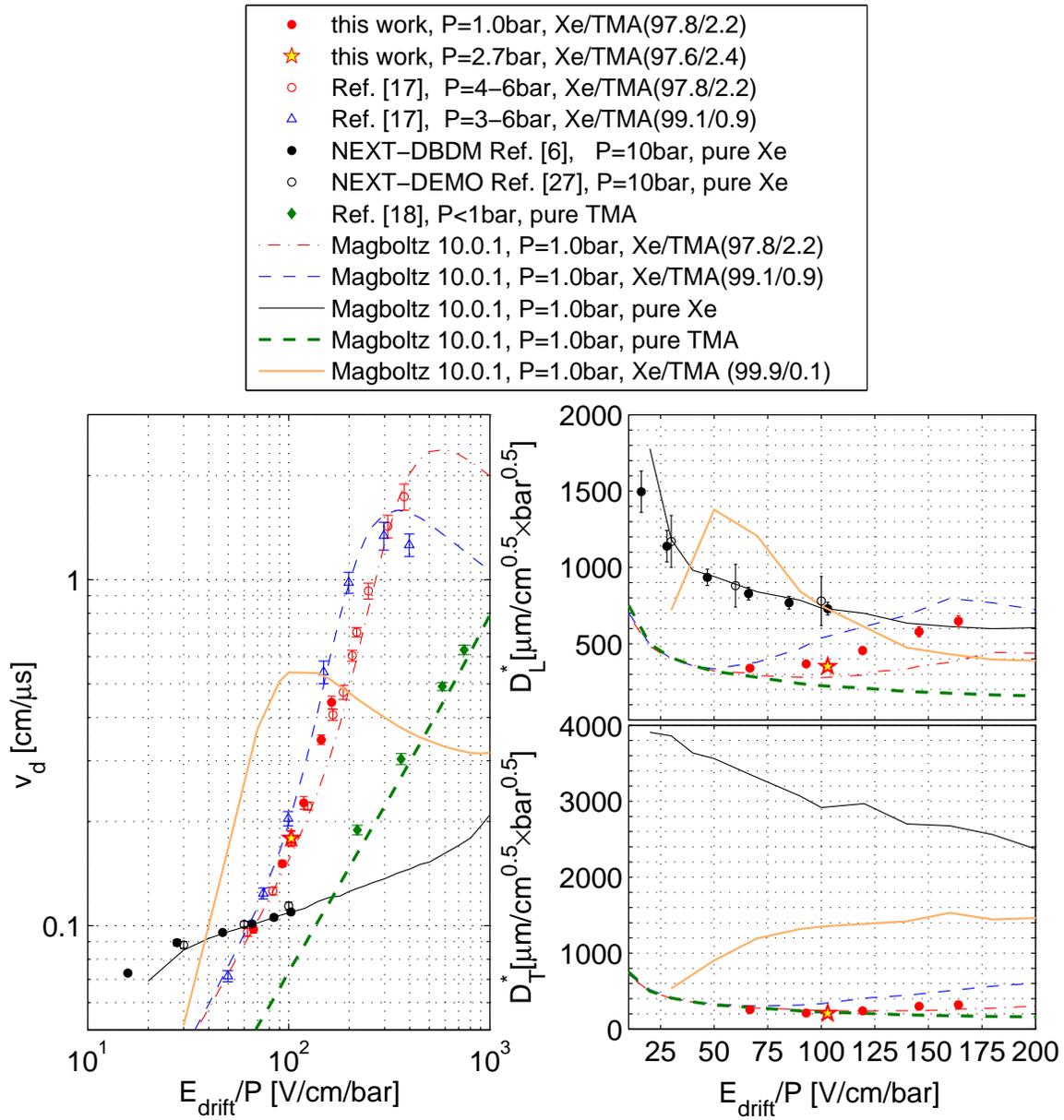}
\caption{Left: world-compilation of drift velocity data for Xe/TMA mixtures including the results from the present work (full red circles, star).(Note: blue triangles have been obtained in an analysis performed after publication of \cite{Diana2}). Right: longitudinal and transverse diffusion, for which only data from pure Xenon and from the present work are available. The estimate of the transverse diffusion is based on the approximate method discussed in section 4, for which only statistical uncertainties are given (smaller than the data points). Magboltz simulation performed at $20\,^\circ$C and 1 atm.}
\label{vdrift}
\end{figure}

The compilation of the drift velocity and longitudinal diffusion coefficient obtained in this work is given in Fig. \ref{vdrift} and table \ref{atta_tab}. For the longitudinal diffusion coefficient the statistical uncertainty of the fit to eq. \ref{diffL} is typically 1\%, to be added to the uncertainty in the drift velocity (5\%) propagated through eqs. \ref{diffL} and \ref{Deff}, and 5\% from the reported difference between different PSA methods, adding to about 5.5\%. We assign a 2\% uncertainty to the estimate of the reduced field (smaller than the data points), arising from the sum in quadrature of the accuracy of the pressure gauge ($\sim\!1$\%) and the present mechanical accuracy in the definition of the drift region ($1.7$\%). The statistical uncertainty in the TMA concentration, obtained from a dedicated calibration of the mass spectrometer with known Xe/TMA proportions, is 2\% (relative, hence $2.20\%\pm0.04\%$ typically). Systematic uncertainties in the calibrating/measuring procedure were assessed by performing measurements of identical admixtures under different conditions
(history, input pressure at the capillary and mixing procedure). The maximum relative deviation found was 10\%, that can be interpreted as the present systematic uncertainty on the TMA estimate.

Fig. \ref{vdrift} includes the latest Magboltz (v10.0.1) microscopic simulations \cite{Magboltz} (performed at $P=1013$\,mbar and $T=20\,^\circ$C), that are seen to provide a reasonable description of all existing Xe/TMA data. The agreement with Magboltz can be further improved by artificially decreasing the TMA concentration by from 2.2\% to 1.4\%, seemingly well beyond the experimental uncertainty of our mass spectrometer. On the other hand, a simultaneous description of $v_d$ and $D_L^*$ within the reported uncertainties does not seem to be possible with Magboltz at the moment even if leaving the TMA concentration free. It must be noted that, in simulation, the overall level of uncertainty of the TMA characterization at low fields is dominated by the (unknown) inelastic cross-section. It is very suggestive that the present best guess for the latter has an estimated uncertainty around 20\% \cite{Biagi} thus at the same level than the reported 30\% effect.

A reduction of the longitudinal diffusion coefficient in Xe/TMA (97.8/2.2) by a factor $2$-$3$ relative to that in pure Xenon can be clearly observed in Fig. \ref{vdrift}-right, besides showing an increasing trend with field (contrary to pure Xenon). According to Magboltz the transverse diffusion may be reduced up to a factor 10 in Xe/TMA mixtures. The increasing behaviour seen in simulation as a function of the electric field as well as the decreasing behaviour with pressure are qualitatively observed in our data (Fig. \ref{Dl_Dt}-right). Fig. \ref{vdrift} shows also for illustration a quantitative estimate of the transverse diffusion as obtained from a simplified analytical description of the behaviour of the magnitude $\epsilon_M/\epsilon_{evt}$ (see next section for details). Notably, and contrary to the general behaviour expected for quenched noble gases, at the intended fields in the NEXT experiment ($E_{drift}/P=25$-$100$\,V/cm/bar) the addition of TMA can modestly reduce the drift velocity relative to that of pure Xenon in 1\%TMA admixtures, while $D_L$ can exceed the values for pure Xenon in 0.1\%TMA ones. Generally, diffusion-wise, the best characteristics of the Xe/TMA mixture in the $E_{drift}/P=25$-$100$\,V/cm/bar range are found for $\sim\!1$\,\%TMA admixtures or beyond.

\begin{figure}[h]
\centering
\includegraphics*[width=10cm]{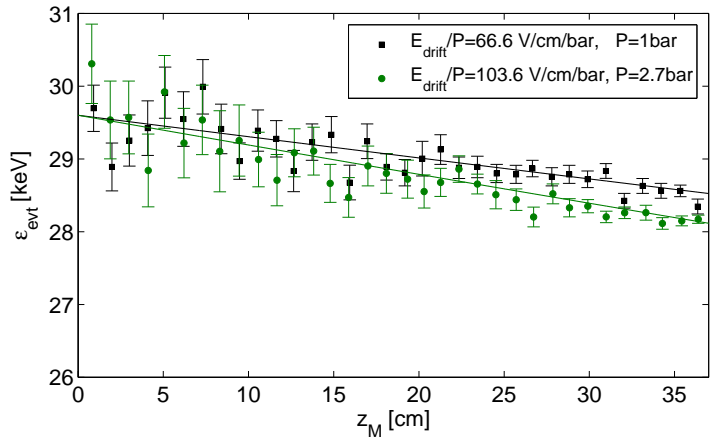}
\caption{Charge loss as a function of the drift distance for typical drit fields at $P=1$\,bar (2.2\% TMA), squares,  and $P=2.7$\,bar (2.4\% TMA), circles. The trend is fitted to an exponential law that allows extracting the attachment coefficient. The data has been up-scaled by $\sim\!5$\% relative to the energy calibration procedure described in the text (that does not use $z$ information at the moment), in order to provide agreement at $z=0$ with the known escape peak energy.}
\label{atta_fig}
\end{figure}

At last, the attachment coefficient of our gas mixture can be obtained from the trends of the event charge/energy as a function of the drift position $z_M$, obtained once random coincidences have been subtracted in order to enhance the sample purity (Fig. \ref{W_t_corr}). Therefore, the $z$-position was sliced and a Gaussian fit in the 30\,keV region performed for each slice, resulting in the data sets shown in Fig. \ref{atta_fig} for two typical cases. An exponential fit yields attachment coefficients of the order of 10\% per meter, showing little dependence with the operating pressure and field. This figure, likely dominated by impurities in the present system, sets an important upper bound to Xe/TMA mixtures whose attachment properties are not known. The (tolerable) 10\% observed effect sets an important benchmark value for the usability of Xe/TMA mixtures in m-scale experiments as NEXT-100. A compilation of the attachment coefficients together with the fit errors is given in table \ref{atta_tab}.


\begin{table}[h]
  \centering
  \begin{tabular}{|c|c|c|c||c|c|c|}
     \hline
     $E/P$[V/cm/bar] & $v_d$[cm/$\mu$s] & $D_L^*$[$\mu$m$/\sqrt{\tn{cm}} \times \sqrt{\tn{bar}} $] & $\eta$[m$^{-1}$] & TMA(\%) & $P$[bar] \\
     \hline
     $66.6\pm1.3$  & $0.097 \pm 0.005$ & $340 \pm 19 $ & $0.10 \pm 0.01$  & 2.2 & 1.0 \\
     \hline
     $93.0\pm1.9$  & $0.151 \pm 0.007$ & $368 \pm 20 $ & $0.08 \pm 0.02$  & 2.2 & 1.0 \\
     \hline
     $119.2\pm2.4$ & $0.227 \pm 0.011$ & $456 \pm 25$  & $0.08 \pm 0.01$  & 2.2 & 1.0 \\
     \hline
     $145.5\pm2.9$ & $0.345 \pm 0.017$ & $579 \pm 32$  & $0.10 \pm 0.01$  & 2.2 & 1.0 \\
     \hline
     $164.0\pm3.3$ & $0.442 \pm 0.022$ & $649 \pm 36$  & $0.07 \pm 0.04$  & 2.2 & 1.0 \\
          \hline
     $103.6\pm2.1$ & $0.179 \pm 0.009$ & $351 \pm 18$  & $0.14 \pm 0.01$  & 2.4 & 2.7  \\
     \hline
   \end{tabular}
  \caption{Table with some parameters for Xe/TMA mixtures extracted from this work. The dominant source of the reported uncertainties is systematic, within the range 2.0-5.5\%, except for $\eta$ where uncertainties proceed from the fit. The relative uncertainty in the TMA concentration is 2\%(sta) and 10\%(sys), the accuracy of the pressure gauge is 10\,mbar. All measurements have been performed at about $20^{\circ}$C with a maximum $5^{\circ}$C deviation during the chamber filling. Although we attribute the observed attachment to impurities, it can be interpreted as an upper bound to the one inherent to the Xe/TMA mixture.}
  \label{atta_tab}
\end{table}

\section{Discussion}\label{discussion}

A full experimental characterization of the dynamics of the primary electron cloud needs of a quantitative estimate of the diffusion coefficient in the direction perpendicular to the electric field, $D_T^*$. It must be noted that according to simulation the transverse diffusion coefficient in present conditions is of the order of $D_T^*=250\,\mu$m$/\sqrt{\tn{cm}} \times \sqrt{\tn{bar}}$ (Fig. \ref{vdrift}, red line). Therefore, the anticipated radial width of a point-like electron cloud is 1.5(0.94)\,mm for 1(2.7)\,bar when bridging the full anode-cathode distance, considerably smaller than the pixel size of the readout plane. On top of this difficulty, the initial size of the ionization cloud stemming from $\sim30\,$keV $\gamma$ interactions needs to be considered, provided it is expected to be also at the mm-scale (table \ref{sources}).

\begin{figure}[h]
 \centering
 \includegraphics*[width=7.5cm]{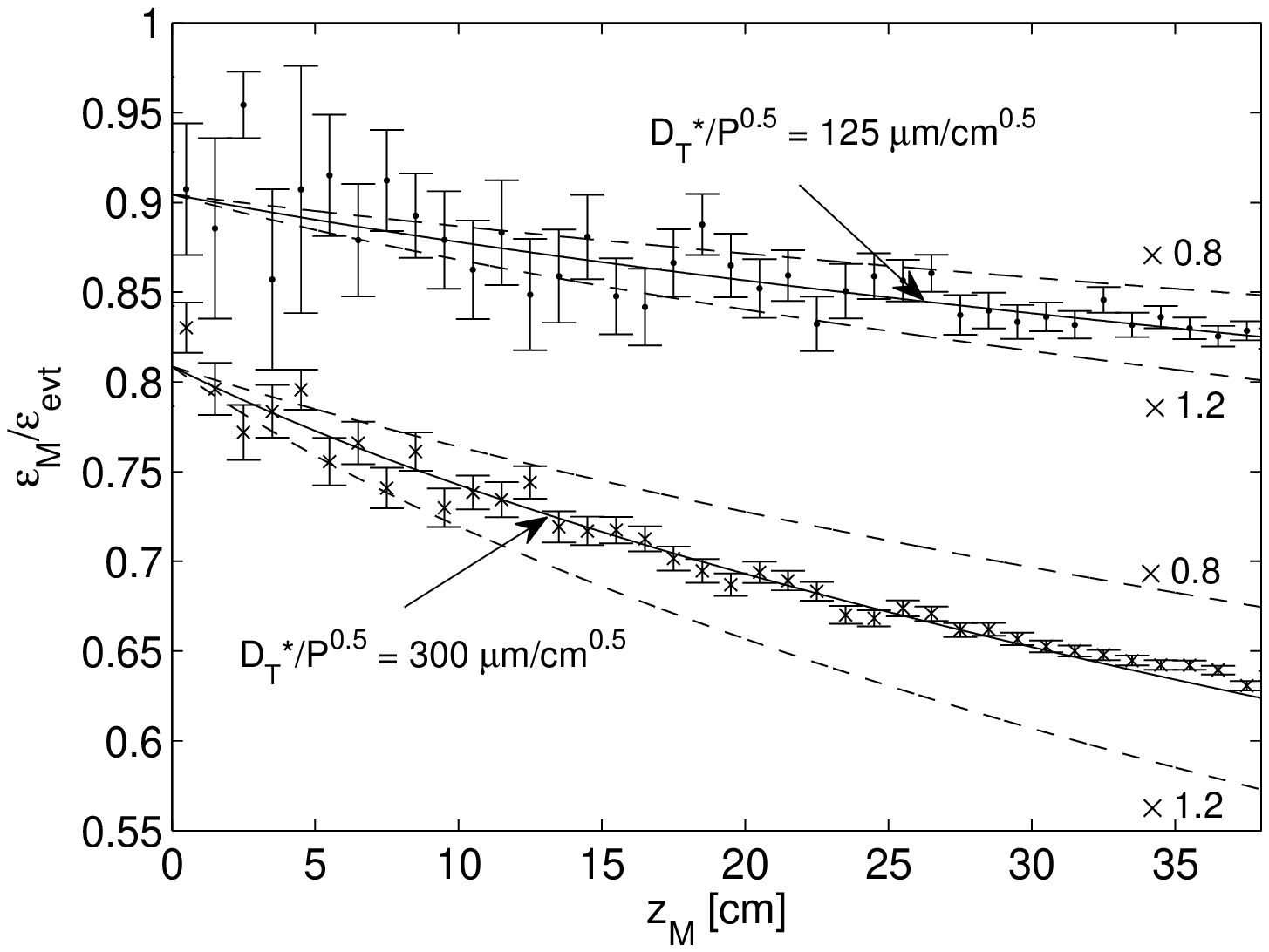}
 \includegraphics*[width=7.5cm]{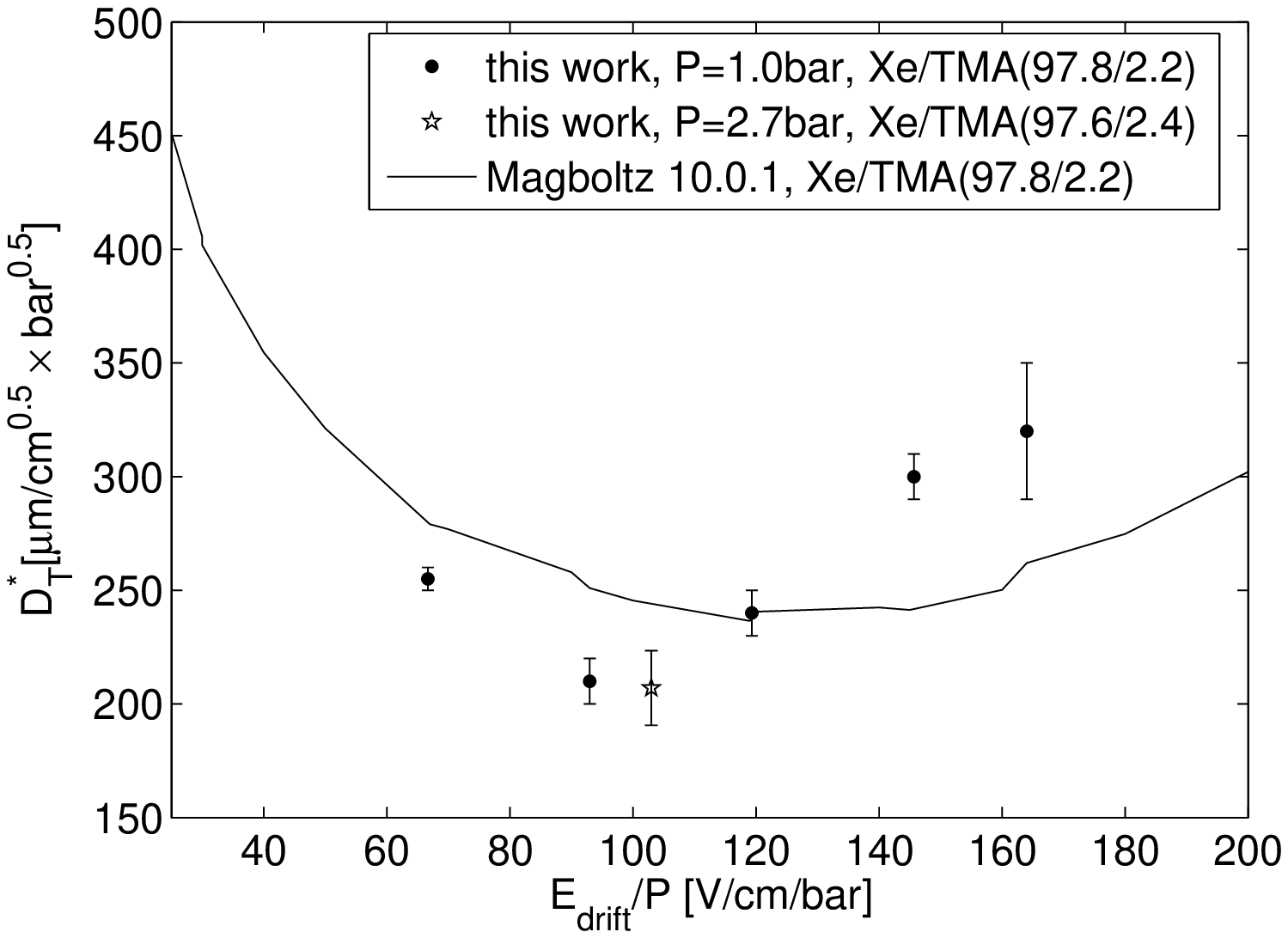}
 \caption{Left: behaviour of the highest charge fraction contained in a single pixel $\varepsilon_M/\varepsilon_{evt}$ as a function of the $z$-position of the deposit for $E_{drift}/P=145.5$\,V/cm/bar at $P=1$\,bar (crosses) and $E_{drift}/P=103.6$\,V/cm/bar at $P=2.7$\,bar (points). The transverse diffusion coefficients extracted from the fit proposed in the text are indicated together with the trends for 20\% variations around it, to illustrate the sensitivity of the method. Right: transverse diffusion coefficient obtained from this analysis including statistical uncertainties only. Systematic effects related to the naive  treatment of the initial ionization cloud have not been evaluated.}
 \label{Dl_Dt2}
 \end{figure}

We tentatively use here a simple method to estimate the transverse diffusion coefficient for cases where the typical charge spread is considerably smaller than the pixel size, a condition that makes a standard analysis based on a barycenter calculation particularly difficult. The main strength of the approach is that it does not rely on simulation and it is based on a simple parameterization of the ratio $\epsilon_M/\epsilon_{evt}$ earlier defined, namely, the highest charge fraction contained in a single pixel as a function of the drift distance. The starting point of the method is the expression for the average behaviour of this observable in the absence of instrumental effects:
\beq
<\frac{\epsilon_M}{\epsilon_{evt}}> = \left[\frac{1}{L}\int_{-L/2}^{L/2} dx \int_{-L/2}^{L/2} \frac{1}{\sqrt{2\pi\hat{v}_r}}~e^{{-\frac{(x-x_0)^2}{2\hat{v}_r}}}dx_0 \right]^2 \label{eq_qovq_1}
\eeq
where $L=8\,$mm is the pixel side and $\hat{v}_r$ stands for the transverse (radial) variance of the cloud:
\beq
\hat{v}_r = {D_T^*}^2 \times \frac{z_M}{P} + \hat{v}_{r,0}
\eeq
The effect introduced by the initial electron cloud is therefore contained in $\hat{v}_{r,0}$. The square in eq. \ref{eq_qovq_1} reflects the fact that the double-integral over the 2 dimensions of the pixel ($x$, $y$) factors out in two equal pieces. The solution to eq. \ref{eq_qovq_1} simply reads:
\beq
<\frac{\epsilon_M}{\epsilon_{evt}}> = \frac{\left[2 \hat{v}_r\left(e^{-\frac{L^2}{2\hat{v}_r}}-1\right) + \sqrt{2\pi}L\sqrt{\hat{v}_r}\tn{erf}\left(\frac{L}{\sqrt{2\hat{v}_r}}\right)\right]^2}{2\pi L^2 \hat{v}_r} \label{QovQ_eq_fit}
\eeq
A 2-parameter fit to the experimental values of $<\frac{\epsilon_M}{\epsilon_{evt}}>$ returns $D_T^*$ and $\hat{v}_{r,0}$. Two exemplary cases for high and low pressure are shown in Fig. \ref{Dl_Dt2}-left for $E_{drift}/P = 103.6$\,V/cm/bar at 2.7\,bar (points), and for $E_{drift}/P = 145.5$\,V/cm/bar at 1\,bar (crosses). The fit returns the following values: $D_T^*/\sqrt{P}=125 \pm 10\,\mu\tn{m}/\sqrt{\tn{cm}}$, $\sqrt{\hat{v}_{r,0}}=0.49 \pm 0.03$\,mm (2.7\,bar) and $D_T^*/\sqrt{P}=300 \pm 10 \,\mu\tn{m}/\sqrt{\tn{cm}}$, $\sqrt{\hat{v}_{r,0}}=1.01 \pm 0.02$ mm (1\,bar). The overall results from the fit to all available data are shown in Fig. \ref{Dl_Dt2}-right with only statistical uncertainties (as derived from the weighted fit) shown, and compared with Magboltz simulations. In line with the previous observations for $v_d$ and $D_L^*$ an approximate agreement is found, improving if slightly increasing the TMA concentration in simulation.

Some caveats are worth being noted at this point: the initial transverse size of the ionization cloud $\hat{v}_{r,0}$ is expected to be independent from the drift field, however up to 20\% deviations are observed in the fitted data, well above the statistical uncertainty. At 1\,bar, this magnitude averaged over all drift fields is $\sqrt{\hat{v}_{r,0}}(1\,\tn{bar})=1.0 \pm 0.1$\,mm, considerably smaller than the csda range (assuming straight propagation) expected from table \ref{sources}. This value is within a factor of 2 of the one extracted at 2.7\,bar, showing a scaling behaviour somewhere in between $1/P$ and $1/\sqrt{P}$, a fact difficult to assess without a detailed microscopic simulation. On another hand, the values for the electron cloud obtained from inspection of the intercepts in the $\sigma_{t,M}-z_{M}$ correlation curve (Fig. \ref{Dl_Dt}-left) slightly differ from this analysis:
\beq
\sqrt{\hat{v}_{z,0}} = \sqrt{\sigma_0^2-\sigma_{0,FEE}^2}\times v_d
\eeq
Taking as an estimate for $\sigma_{0,FEE}=1.15\,\mu$s (the smallest intercept observed in data) and using the $\sigma_0$ measured at the lowest field returns $\sqrt{\hat{v}_{z,0}}(1\,\tn{bar})=0.6$\,mm, about 40\% smaller than the result from the transverse analysis. Despite these potential biasing effects not being accounted for, the present analysis seems to soundly set a scale for the transverse coefficient $D_T^*\sim250$-$300\,\mu\tn{m}/\sqrt{\tn{cm}} \times \sqrt{\tn{bar}}$ in the reduced field range $E_{drift}/P=50$-$200$\,V/cm/bar for $\sim2$\,\%TMA admixtures.

A new simulation software (Degrad) has been recently developed by S. Biagi in order to calculate the electron clouds stemming from X-ray interactions in gases \cite{Magboltz}, so in principle it should be possible to accurately evaluate the proposed method against simulation in the near future, as well as using bench-marking reference data to allow for a more accurate determination of $D_T^*$.

\section{Conclusions}\label{Conclusions}
Several system aspects of a 70\,l high pressure Xenon TPC operated with a microbulk Micromegas readout plane have been presented. The TPC was operated under low energy $\gamma$ rays with energies around the characteristic Xenon $K_{\alpha,\beta}$ emission (30\,keV), collecting about $2.4\times 10^6$ events. Apart from a temporary technical problem affecting one quadrant, the TPC achieved an energy resolution after calibration of $10.6$\%\,FWHM(1\,bar) and $12.5$\%\,FWHM(2.7\,bar) over its entire active volume ($\sim25$\,l). Inter-quadrant gain variations around 10\% were observed while the inter-pixel gain spread stayed within 20\%, underlining the excellent uniformity of response of this type of readouts when aiming at large area coverage ($700\,$cm$^2$ in this case). No special issues related to the pressure increase were observed except for a 25\% reduction of the workable gain at 2.7\,bar ($m=1.6\times10^3$). The chamber has been conceived for operation under high energy $\gamma$'s at 10\,bar, a step that is currently being preceded by an upgrade of the gas system.

For the first time the longitudinal diffusion coefficient in Xe/TMA mixtures has been extracted, in particular for a 2.2\% TMA admixture. A reduction of its value by about $\times 2$-$3$ with respect to pure Xenon was observed in the 50-100\,V/cm/bar reduced drift field regime, roughly as expected from microscopic modeling (Magboltz v10.0.1). On the other hand, the qualitative behaviour observed for the transverse diffusion coefficient agrees with the one obtained from Magboltz (that anticipates a reduction by a factor 10) and indeed a value at the $\sim250$-$300\,\mu\tn{m}/\sqrt{\tn{cm}} \times \sqrt{\tn{bar}}$ level predicted by simulation is hinted by present data. A significant improvement of the 2-blob topological signature in the $\beta\beta0$ decay of $^{136}$Xe is not guaranteed by the reduced diffusion and requires further studies, but its potential makes the case appealing. Generally, diffusion-wise, the best characteristics for Xe/TMA mixtures in the $E_{drift}/P=25$-$100$ V/cm/bar range are found for $\sim\!1$\% TMA admixtures or beyond, however it is necessary to verify that such TMA concentrations are compatible with the fluorescence yields and Penning transfer rates required for $0\nu\beta\beta$ searches, and this is currently under investigation.

The observed attachment coefficient was little dependent on pressure and electric field, showing values around 10\% for a 1 meter electron drift (correspondingly, $10.0\pm1.1\,$ms electron life-time for the lowest reduced field studied $E_{drift}/P=66.6$ V/cm/bar). This level is presumably dominated by impurities in the gas and hence it should be interpreted as an upper bound to the attachment inherently coming from the Xe/TMA mixture. In the seemingly unlikely case that such an upper bound would be saturated, a 1\,m scale experiment (as NEXT-100) could still make use of it at a very mild charge loss if Xe/TMA finally proves to be a golden combination for this type of physics.

\section*{Acknowledgements}

NEXT is supported by the following agencies and institutions: the Ministerio de Econom\'ia
y Competitividad (MINECO) of Spain under grants CONSOLIDER-Ingenio 2010 CSD2008-0037 (CUP),
FPA2009-13697-C04-04 and FIS2012-37947-C04; the Director, Office of Science, Office of Basic
Energy Sciences, of the US Department of Energy under contract no. DE-AC02-05CH11231; and
the Portuguese FCT and FEDER through the program COMPETE, projects PTDC/FIS/103860/2008
and PTDC/FIS/112272/2009. J. Renner (LBNL) acknowledges the support of a US DOE NNSA
Stewardship Science Graduate Fellowship under contract no. DE-FC52-08NA28752.

The development of NEXT-MM in particular has important contributions from the European Research Council T-REX Starting Grant ref. ERC-2009-StG-240054 of the IDEAS program of the 7th EU Framework
Program; and also MINECO under grants FPA2008-03456, FPA2011-24058 and CPAN under grant CSD2007- 00042.

This work would not have been possible without the encouragement of our RD51 colleages and the dedicated effort of R. de Oliveira and his team at CERN to provide these unique detectors. Discussions with O. Sahin, R. Veenhof and S. Biagi are greatly acknowledged.

\end{document}